\pdfoutput=1

\documentclass[11pt]{article}

\usepackage[preprint]{acl}

\usepackage{times}
\usepackage{latexsym}

\usepackage[T1]{fontenc}

\usepackage[utf8]{inputenc}

\usepackage{microtype}

\usepackage{inconsolata}

\usepackage{graphicx}

\usepackage[utf8]{inputenc} 
\usepackage[T1]{fontenc}    
\usepackage{hyperref}       
\usepackage{url}            
\usepackage{booktabs}       
\usepackage{amsfonts}       
\usepackage{nicefrac}       
\usepackage{microtype}      
\usepackage{xcolor}         
\usepackage{wrapfig}
\usepackage{graphicx} 
\usepackage{booktabs} 
\usepackage{arydshln} 
\usepackage{multirow} 
\usepackage{float} 
\usepackage{subcaption}
\usepackage{dashrule}
\usepackage[normalem]{ulem}
\usepackage{enumitem}
\usepackage{xspace}
\usepackage{paralist}
\usepackage{todonotes}
\newcommand{\schemelong}{\textsc{LibEvolutionEval}\xspace}

\usepackage{titlesec}
\titlespacing{\paragraph}{%
  0pt}{
  0.1\baselineskip}{
  1em}%

\title{\schemelong: A Benchmark and Study \\ for Version-Specific Code Generation}

\author{
  \texttt  Sachit Kuhar$^{1}$
  ~~
  Wasi Uddin Ahmad$^{2*}$
  ~~
  Zijian Wang$^{1}$
  ~~ 
  Nihal Jain$^{1}$ 
  ~~
  Haifeng Qian$^2$\thanks{~Work done at AWS AI Labs. $^{1}$AWS AI Labs. $^{2}$NVIDIA.  Correspondence: \texttt{\{skuhar, rabaisha\}@amazon.com}} \\
  \textbf{Baishakhi Ray}$^{1}$ 
  ~~
  \textbf{Murali Krishna Ramanathan}$^{1}$ 
  ~~
  \textbf{Xiaofei Ma}$^{1}$
  ~~
  \textbf{Anoop Deoras}$^{1}$ 
  }

\begin{document}
\maketitle
\vspace{-5pt}
\begin{abstract}
\vspace{-5pt}
Recent advancements in code completion models have primarily focused on local file contexts~\citep{ding2023crosscodeeval, jimenez2024swebench}. However, these studies do not fully capture the complexity of real-world software development, which often requires the use of \textbf{rapidly-evolving} public libraries.  To fill the gap, we introduce \schemelong, a detailed study requiring an understanding of library evolution to perform in-line code completion accurately. \schemelong provides a version-specific code-completion task comprised of eight libraries (\texttt{torch, torchvision, scipy, pil, tqdm, pyyaml, matplotlib}, and \texttt{pandas}) as they evolve over the year along with a detailed analysis of the evolution of two popular and well-maintained public libraries: \texttt{PyTorch} and \texttt{Matplotlib}. 
We evaluate popular public models and find that public library evolution significantly influences model performance. 
We explored mitigation methods by studying how retrieved version-specific library documentation and prompting can improve the model's capability in handling these fast-evolving packages, paving a promising future path in better handling fast-evolving libraries. 
\looseness=-1
\end{abstract}

\section{Introduction}

Large Language Models for code (\emph{a.k.a.} code LLMs)~\citep{li2023starcoder, lozhkov2024starcoder, roziere2023code} have significantly advanced developer productivity through improved code completion tasks. 
These models are pivotal not only in code completion, but also in debugging, code summarization, and language translation for software development \citep{yan2023codetransocean, lachaux2020unsupervised, roziere2021leveraging, min2023beyond}.
These models are usually evaluated either with code contest dataset~\citep{doi:10.1126/science.abq1158} or 
with a focus on local files for context to enhance the completion of the function~\citep{chen2021evaluating, ding2023static, athiwaratkun2022multi, ding2023crosscodeeval, jimenez2024swebench}. 
However, these studies do not fully encompass the complexities of real-world software development, which requires public libraries.
Complexity of code completion with public library APIs increases, as the APIs often evolve---some APIs change their signature, some gets deprecated, while many new APIs surfaced in this evaluation process~\citep{mcdonnell2013empirical}.
While some works perform code completion involving public libraries~\citep{liao2023context, zan2022cert}, use documentation of the library for prediction~\citep{qin2023toolllm}, and show that {zero shot} code completions suffer from hallucinations~\citep{patil2023gorilla}, these works do not focus on the rapidly evolving nature of public libraries.

\begin{figure*}
    \centering
    \includegraphics[width=0.84\textwidth]{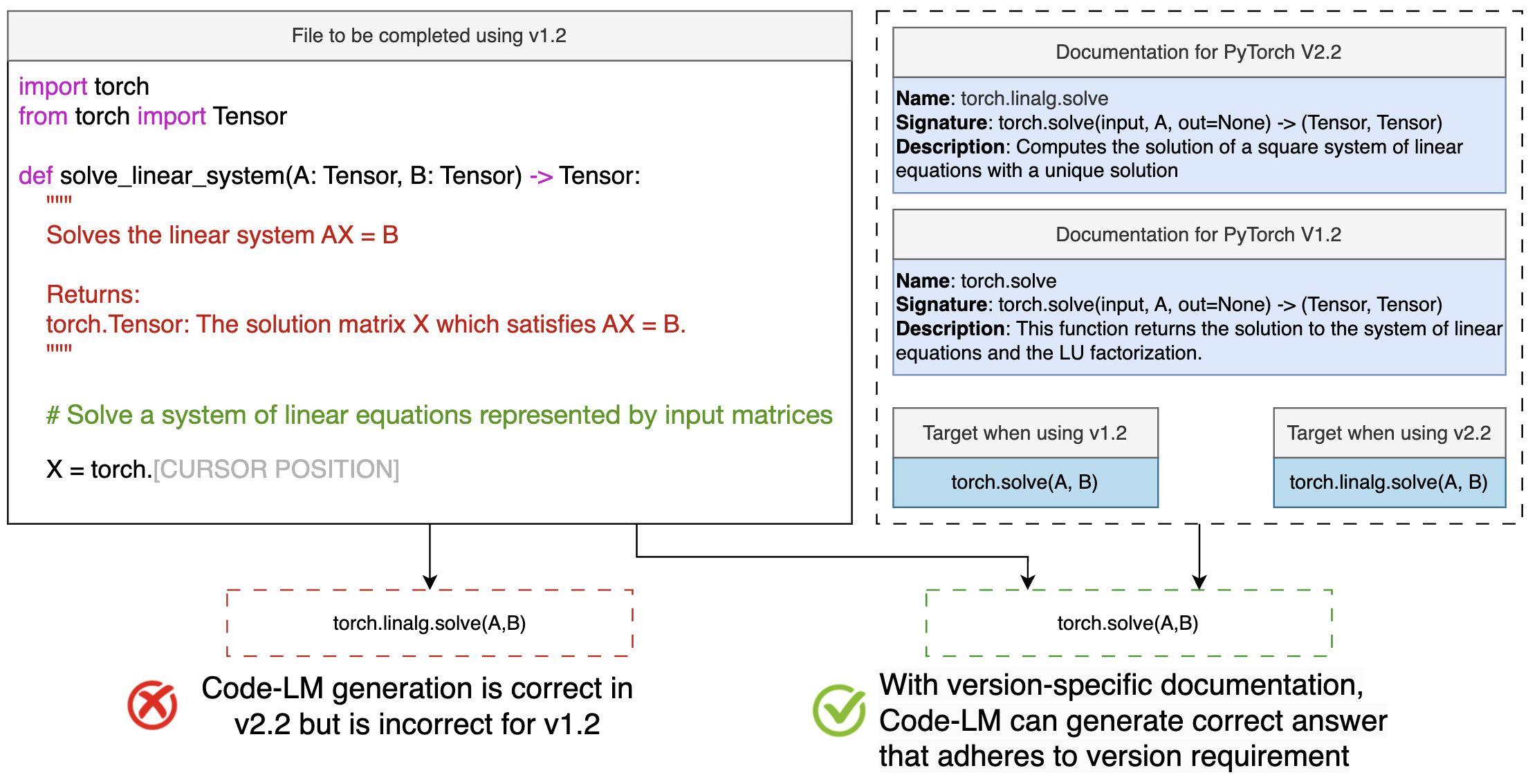}
    \vspace{-4pt}
    \caption{An example of a code completion scenario under \schemelong. The incomplete code snippet on the left requires the correct API method to solve a linear system specified by two PyTorch tensors. The code LLM performs incorrect code completions due to version mismatch. 
    The version-specific documentation is a potential {augmentations} that can assist the LLM to perform correct and version-dependent completion.
    }
    \vspace{-8pt}
    \label{fig:context_example}
\end{figure*}

Large Language Models are trained on extensive corpora of open-source code, which likely includes public libraries. 
Consequently, while LLM-generated code may appear reasonable, it might not be accurate for the specific version of the library being used, leading to version-dependent performance issues.
Figure~\ref{fig:context_example} shows that Code LLM's generation is correct for \texttt{v2.2} but incorrect for \texttt{v1.2}.   
This variability is significant since developers often work with different library versions -- newest versions for current projects and older ones for legacy code maintenance. Therefore, the developer's experience with coding assistants depending on LLMs for code completion can vary greatly depending on their specific use case.

Existing benchmarks and studies do not fully capture evolution, revealing a gap in our current evaluation and understanding of code LLMs. This work focuses on the following research questions: (1) Does the performance of code LLMs change as the library evolves? (2) If yes, can retrieving version-specific meta-data like library documentation mitigate the impact of library evolution on code completion? (3) With the evolution of libraries, new relationships between APIs are introduced, while existing ones are altered. Can code LLMs effectively adapt to these evolving relationships between APIs? (4) Does the introduction, modification, and deprecation of APIs as libraries evolve to make it more challenging for code LLMs to perform accurate code completions?

To investigate these research questions, we introduce LibEvolutionEval, a benchmark and detailed study specifically designed to understand the impact of public library evolution on code completion. LibEvolutionEval offers version-specific code-completion tasks spanning multiple years for eight libraries -- \texttt{torch, torchvision, scipy, pil, tqdm, pyyaml, matplotlib}, and \texttt{pandas}.

LibEvolutionEval also performs a detailed analysis of the evolution of two popular libraries (\texttt{torch and matplotlib}) by providing version-specific meta-data, documentation retrieval tasks, and code-completion tasks. 
It requires code LLMs to perform version-specific code completions under both realistic (where the evaluation examples are sampled from permissively licensed GitHub repositories) and controlled scenarios (where a template uses API documentation to create evaluation examples). It provides version-specific API documentation to investigate the impact of library evolution on embedding models during retrieval.  
It offers tasks based on completion type to compare: (1) completions guided by import statements and clear library prefixes with (2) completions that are object-oriented references and do not have a library-defined prefix. We call them \emph{direct} and \emph{indirect} code completions (see Figure~\ref{fig:api_classification_direct_indirect}), and such tasks evaluate models' ability to adapt to evolving relationships between APIs. Furthermore, LibEvolutionEval also offers tasks based on granularity that compares overall developer experience with performance on specific APIs that have been newly introduced, modified, or deprecated as the library evolves, providing us with insights on the impact of evolution against overall code-completion performance.

\begin{figure*}[th]
    \centering
    \includegraphics[width=0.95\textwidth]{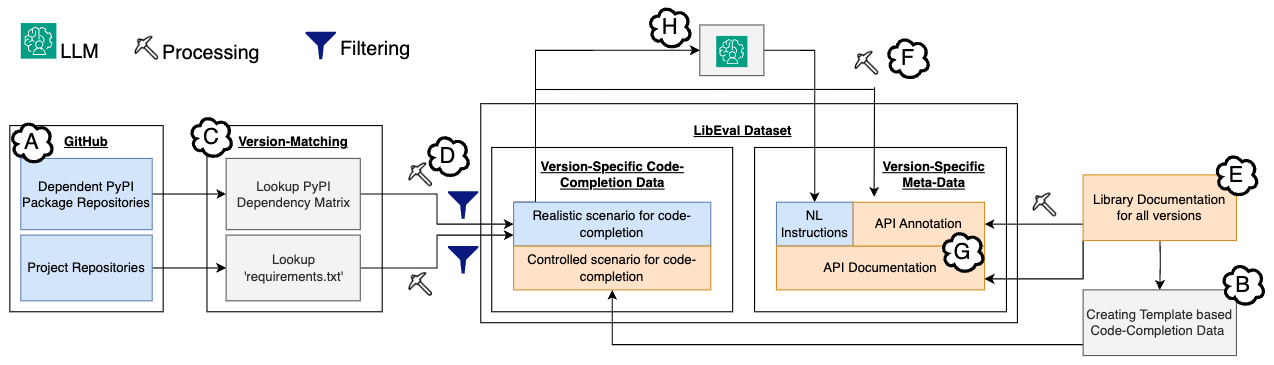}
    \caption{\schemelong's preprocessing pipeline to obtain version-specific code-completions meta-data including documentation, NL instructions, and API annotation.}
    \vspace{-4pt}
    \label{fig:data_pipeline}
\end{figure*}

We conducted a comprehensive evaluation using widely used code LLMs~\citep{lozhkov2024starcoder, jiang2023mistral, openai2024gpt4o} and embedding models~\citep{zhang2024codesage, openai-embedding-ada-002} to report the following insights.
\begin{compactitem}
    \item Code LLMs and embedding models exhibit substantial performance variation as public libraries evolve. Providing version-specific API documentation as a context improves code completion performance but does not entirely address inherent version-based bias in version-specific code completions.
    \item Code LLMs perform indirect API completions better than direct API completions, demonstrating an understanding of evolving relationships between APIs.
    \item Introduction, modification, and deprecation of APIs make it harder for code LLMs to perform code completion where new models might forget old deprecated APIs while old models cannot predict the latest APIs.
\end{compactitem}

\section{\schemelong: Version-Specific Code Completions}
\label{sec:globalimportseval}

Each code completion example in \schemelong~ consists of code prompts ending at a position where the LLM is tasked to complete the missing expression, typically involving one or more API calls to the public library under consideration as shown in Figure~\ref{fig:context_example}. The uniqueness of this dataset lies in its emphasis on version-specific API usage, reflecting scenarios where developers use LLMs to perform code completions for different versions of the same library. We evaluate code completions under two scenarios: \emph{realistic} (GitHub based) and \emph{controlled} (documentation based).

\subsection{Version-Specific Evaluation Creation}
\label{subsec:dataset_collection}

\paragraph{API Usage Collection}\label{para:api_usage_collection}
For a realistic scenario, we focus on data written by real-world developers, specifically from permissively licensed GitHub repositories (Figure~\ref{fig:data_pipeline}-(A) and Figure~\ref{fig:torch_code_example} in the appendix).
This allows us to understand if the impact of API evolution is significant with unknown confounding variables present in real-world code.

For detailed ablations on the other hand, we also simulate a controlled scenario by creating synthetic data for \texttt{Matplotlib} and \texttt{PyTorch} by taking API documentation and converting it to evaluation examples using a template. The template is designed to make the code LLM predict the API name given its description, service name, and mandatory arguments in the left context (Figure~\ref{fig:data_pipeline}-(B) and Figure~\ref{fig:aws_api_template_example} in the appendix).
This allows us to isolate the impact of API evolution without the confounding variables found in real-world code, such as variations in coding styles.

\paragraph{Versioning of API Usage}
For a realistic setting, the GitHub repositories have a ‘requirements.txt’ file that mentions the exact version of the library used to develop it. Additionally, if the GitHub repository is a PyPI package (like \texttt{torchvision}) that depends on the library under consideration (like \texttt{torch}), we use the dependency matrix between different packages to match API usage with the library version (Figure~\ref{fig:data_pipeline}-(C)).
Next, since the data is created from API documentation for the controlled setting, the version of the API usage example is the same as that of the documentation from which it is derived. 

\paragraph{API Evaluation Example Creation}
To ensure the quality of our proposed benchmark, we employ a series of rule-based and model-based post-processing filters (Figure~\ref{fig:data_pipeline}-(D)). 
We typically restrict the left context provided to the model to the scope of the API being completed, limiting it to the class containing the API call.
If no class is present, we include the entire preceding context of the line.
Import statements of the target library are also included to provide the LLM with contextual clues (see \S\ref{sec:data_collection} in appendix). 
Additionally, the initial regex of the API expression (e.g., \texttt{torch} from \texttt{torch.solve()} as illustrated in Figure~\ref{fig:context_example}) is placed just before the cursor position to encourage the model to complete the API expression correctly. Comments are removed from contexts to minimize the risk of API leakage.

\paragraph{API Evaluation Example Selection}
When encountering multiple API calls on the same line (e.g., \texttt{x = torch.ones(x) + torch.zeros(y)}), if an API (e.g., \texttt{torch.ones}) has already been included in the evaluation dataset from this line, subsequent APIs on that line (e.g., \texttt{torch.zeros}) are excluded. This approach ensures diversity in the contexts represented in the dataset. Additionally, we discard examples if the corresponding API call already exists within the collected evaluation dataset for the same source (e.g. GitHub repository). This means that if we have already included an example of a specific API (e.g., \texttt{torch.ones}) from a particular source, we will exclude any additional examples from that same source that use the same API (i.e., \texttt{torch.ones}). This strategy aims to ensure the diversity of API calls within the evaluation dataset.

\paragraph{Documentation Collection}
We systematically collect documentation for detailed analysis of library evolution from their publicly available websites (Figure~\ref{fig:data_pipeline}-(E)). This includes comprehensive details such as API signatures, names, types, input parameters (noting it's optional/mandatory nature), and code usage examples. This documentation serves as a foundation for understanding these APIs' expected usage and evolution over time.

\begin{figure}[t]
\centering
\includegraphics[width=0.45\textwidth]{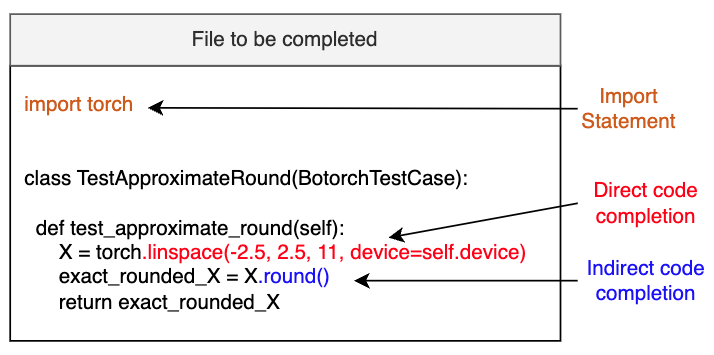}
\vspace{-2pt}
\caption{
APIs classification based on \emph{completion type}.
}
\vspace{-2pt}
\label{fig:api_classification_direct_indirect}
\end{figure}

\begin{figure*}[ht]
    \centering
    \begin{subfigure}[b]{0.49\textwidth}
        \includegraphics[width=\textwidth]{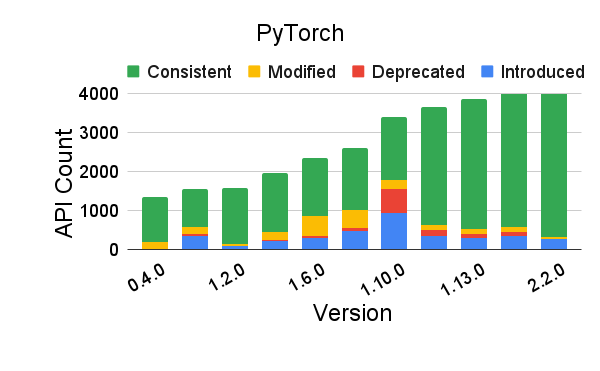}
    \end{subfigure}
    \hfill
    \begin{subfigure}[b]{0.49\textwidth}
        \includegraphics[width=\textwidth]{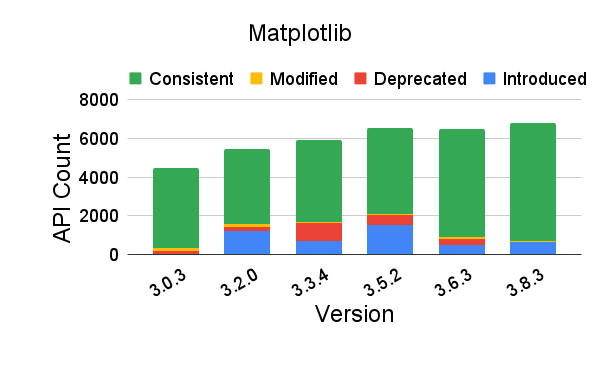}
    \end{subfigure}
    \vspace{-16pt}
    \caption{Illustration of the evolution of PyTorch and Matplotlib public libraries over time. This highlights the rapid evolution of modern public libraries.}
    \label{fig:documentation_visualize}
    \vspace{-4pt}
\end{figure*}

\vspace{-5pt}

\subsection{API Data Classification}
\label{subsec:api_classification}

\paragraph{Completion Type}
This classification assesses an LLM's ability to track the evolving relationships between APIs as a library changes. It does so by comparing completions based on import-driven prefixes with those using open-vocabulary prefixes (Figure~\ref{fig:api_classification_direct_indirect} and Figure~\ref{fig:data_pipeline}-(F)).
These are:
\begin{compactitem}
    \item \textbf{Direct Code Completions}: These completions are driven by import statements, with prefixes derived directly from the public library's import statements. As an example, \texttt{nn.ReLU()} is a direct API completion from \texttt{import torch.nn as nn}. 
    \item \textbf{Indirect Code Completions}: These completions lack a well-defined prefix which originates from referenced objects instantiated through direct API calls. Figure \ref{fig:api_classification_direct_indirect} shows that variable \texttt{X} is defined by \texttt{X = nn.linspace} and is later used in \texttt{X.round}. These completions test a model's deeper contextual understanding, requiring it to identify the corresponding direct API call and comprehend the library's version-specific relationships between APIs to achieve accurate code completion.
\end{compactitem}

\paragraph{Granularity}
This evaluation examines LLMs' ability to adapt to rapid API changes, including introductions, deprecations, and modifications as the library evolves. We annotate APIs using their documentation as (Table~\ref{tab:api_changes2} and Figure~\ref{fig:data_pipeline}-(G)):
\begin{compactitem}
    \item \textbf{Introduced}: API added in the current version and not present in the previous version.
    \item \textbf{Deprecated}: API present in the current version but removed in the next version.
    \item \textbf{Modified}: The name of the API does not change but its arguments are updated.
    \item \textbf{Unchanged}: API does not change when compared to the previous/next version.
\end{compactitem}

We then cross-reference these annotated APIs with the version-specific evaluation examples. This allows us to label both API documentation (see Figure~\ref{fig:documentation_visualize}) evaluation examples based on granularity to create subsets for analysis.

\begin{table}[t]
    \centering
    \resizebox{\columnwidth}{!}{%
    \setlength{\tabcolsep}{2pt}
    \begin{tabular}{l  c c c}
        \hline
        \textbf{Type} & \textbf{Old Version} & \textbf{Current Version} & \textbf{Next Version} \\ \hline
        Introduced API       & Not Supported        & API.Foo(x)               & API.Foo(x)            \\ \hline
        Deprecated API & API.Foo(x)           & API.Foo(x)               & Not Supported         \\ \hline
        Modified API  & API.Foo(x, y)         & API.Foo(x)               &        \\ \hline
        Modified API  &          & API.Foo(x)               & API.Foo(x, y)         \\ \hline
    \end{tabular}%
    }
    \vspace{-2pt}
    \caption{Classification of APIs based on \emph{granularity}.}
    \vspace{-4pt}
    \label{tab:api_changes2}
\end{table}

\subsection{LLM Context Classification}\label{subsec:api_context}

\paragraph{In-File Context} We assess the models' code completion capabilities using only the context available within the current file, replicating a typical development environment scenario. The import statements, the right context, and the left context extracted are given to the model (see Figures~\ref{fig:context_example} and~\ref{fig:prompt_zero_shot}). This methodology ensures that our evaluation accurately reflects the practical conditions faced by developers that use version-unaware code completions and is aimed to serve as a baseline.

\paragraph{Library Version-Aware Context} While the in-file context mimics a realistic code-completion setting 
, there still exists ambiguity for the LLM to perform code completion. For example, there might be two valid responses based on the in-file context corresponding to two different versions of the library, as shown in Figure~\ref{fig:context_example}. To mitigate this issue, we add a comment before left context that tells the LLM the version of the library under consideration.

\paragraph{Version-Specific Retrieved API-Context} 
Development on the success of retrieve-and-generate frameworks for repository-level code completions \citep{zhang2023repocoder, ding2023crosscodeeval}, we adapt this retrieve-and-generate approach for the retrieval of public library documentation. Our documentation retrieval database is organized by versions of public libraries. 
It contains comprehensive metadata, including API signatures, input parameters, usage examples, and detailed natural language descriptions of APIs and their parameters. For each code completion task, we generate a query using the natural language instruction describing the developer's intent. This is done by giving target code completion to Anthropic's Claude v2~\citep{claude} and asking it to not reveal the regular expressions corresponding to the name and input arguments of target code completion (refer to \S~\ref{sec:claude_prompt} for the Appendix and Figure~\ref{fig:data_pipeline}-(H)). We utilize an embedding model (CodeSage~\citep{zhang2024codesage} by default) to determine the similarity between the query and the available API entries, selecting the top 3 matching APIs.
These are then formatted as commented code, incorporating API signatures and parameters, placed before the left context to serve as the version-specific documentation as shown in Figure~\ref{fig:prompt_nl_doc}.\looseness=-1

\subsection{Dataset Statistics and Scope}
\label{sec:data_stats}

\paragraph{Statistics} 
We present the statistics of \schemelong in Table~\ref{tab:data_stat}. We use the StarCoder tokenizer \citep{li2023starcoder} to compute the number of tokens. For version-specific characteristics, see \S\ref{sec:version_level_stats} in the appendix.

\begin{table}[t]
\centering
\resizebox{\linewidth}{!}{
\setlength{\tabcolsep}{2pt}
\begin{tabular}{l r r r}
\toprule
Feature & Assorted & PyTorch & Matplotlib \\
\midrule 
\# API Documentations & - & 29.4K & 35.6K \\
\# Eval Examples & 4.5K & 20.1K & 10.1K \\
Avg. \# lines in prompt & 66.25 & 104.91 & 84.36 \\
Avg. \# tokens in prompt & 732.06 & 1149.34 & 995.91 \\
Avg. \# lines in reference & 1.27 & 1.21 & 1.25 \\
Avg. \# tokens in reference & 18.74 & 13.73 & 17.47 \\
\bottomrule
\end{tabular}
}
\vspace{-4pt}
\caption{\schemelong~statistics.}
\vspace{-4pt}
\label{tab:data_stat}
\end{table}

\paragraph{Scope} 
In addition to left contexts and target code completions, we include the subsequent code lines from the source code files in \schemelong examples.
By providing the source code lines both to the left (prompt or prefix) and to the right (suffix) of the references,  \schemelong enables the evaluation of code LLMs for their fill-in-the-middle (FIM) capabilities \citep{bavarian2022efficient}. 
Furthermore, the meta-data from documentation allows us to conduct evaluations using RAG.

\section{Experimental Setup}
\label{sec:experiments}

\begin{figure*}[ht]
    \centering
    \includegraphics[width=0.85\textwidth]{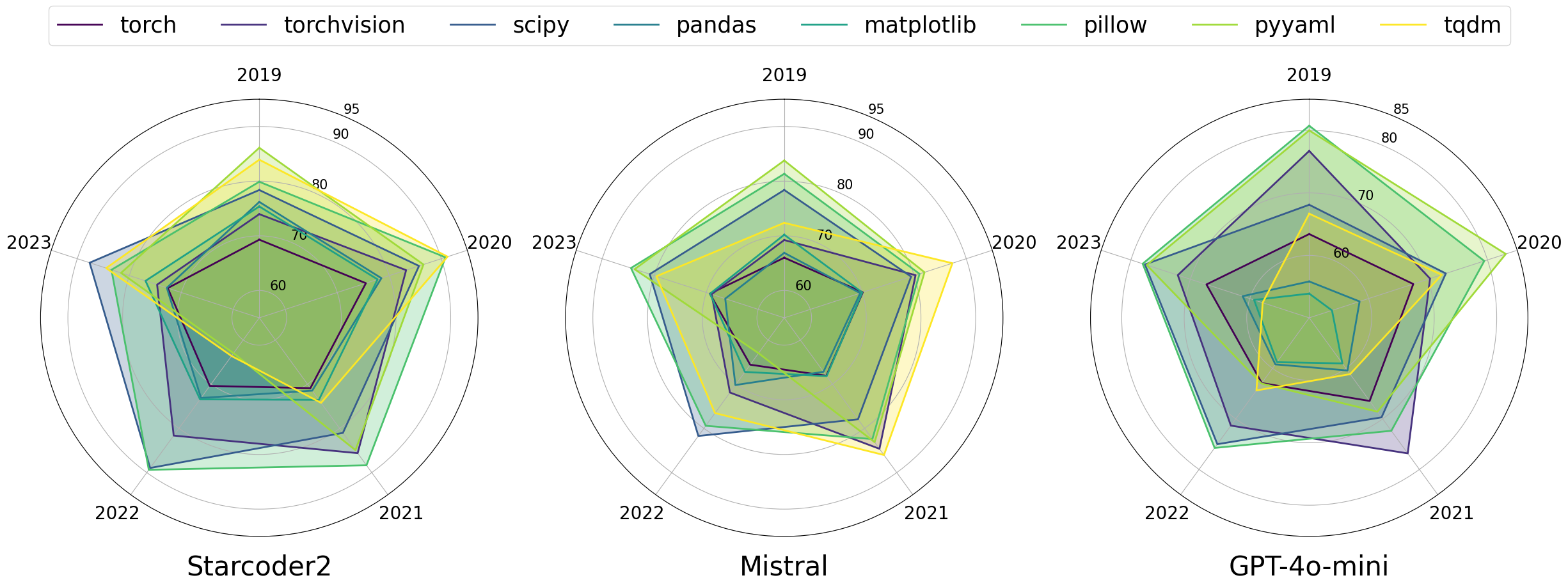}
    \vspace{-2pt}
    \caption{
    Illustration of the code completion performance of the Starcoder2, Mistral, and GPT-4o-mini models by measuring the F1 score. The performance of code LLMs varies significantly as libraries evolve.
    }
    \vspace{-2pt}
    \label{fig:zero_shot_eval}
\end{figure*}

\begin{table*}[ht!]
    \centering
    \resizebox{0.94\textwidth}{!}{
    \begin{tabular}{|l|l|l|l|l|}
        \hline
        \textbf{Model} & \textbf{Completion Strategy} & \textbf{Context Setting} & \textbf{PyTorch} & \textbf{Matplotlib} \\ \hline
        \multirow{3}{*}{Starcoder2-7B} &  & In-File (Not Version-Aware) & 68.8 & 69.7 \\
        & Fill-in-the-Middle & + Version-Aware  & 69.3 & 70.1 \\
        &  & + Version-Aware RAG  & \textbf{73.3} & \textbf{75.4} \\ \hline
        \multirow{3}{*}{Mistral-7B} &  & In-File (Not Version-Aware) & 65.8 & 60.18 \\
        & Left-Context Only & + Version-Aware  & 66.04 & 61.2 \\
        &  & + Version-Aware RAG  & \textbf{67.6} & \textbf{69.05} \\ \hline
        \multirow{3}{*}{GPT-4o-Mini} &  & In-File (Not Version-Aware) & 64.3 & 52.5 \\
        & Instruction-based (w/ Example) & + Version-Aware  & 64.78 & 53.1 \\
        & & + Version-Aware RAG  & \textbf{70.14} & \textbf{66.7} \\ \hline
    \end{tabular}
    }
    \caption{Code completion performances under different input types and context prompting strategies. Each model is evaluated in three context settings: in-file (not version-aware), version-aware, and version-aware RAG.}
    \vspace{-4pt}
    \label{tab:prompting_table}
\end{table*}

\paragraph{Models}
\label{subsec:models}
We benchmark public code LLMs: Mistral~\citep{jiang2023mistral}, StarCoder2~\citep{lozhkov2024starcoder}, GPT-4o-mini~\citep{openai2024gpt4o} and CodeGen 1.0~\citep{Nijkamp2022ACP}. We benchmark version-specific retrieval tasks using CodeSage~\citep{zhang2024codesage} and OpenAI-ada-002~\citep{openai-embedding-ada-002}. Lastly, we conducted scaling experiments with StarCoder~\citep{li2023starcoder} (1B, 3B, 7B), StarCoder2 (3B, 7B, 15B), and CodeSage (Small, Large).

\paragraph{Evaluation Metrics}
For code completion, we concentrate on the correctness of APIs called by calculating the F1 score \citep{ding2023crosscodeeval}. For documentation retrieval, to evaluate the performance of embedding models by using Mean Reciprocal Rank (MRR), assessing how well they retrieve version-specific documentation and whether their performance varies with library evolution.

\paragraph{Inference}
We maintain uniform hyperparameters across all models. The maximum sequence length is 8K tokens, with each context trimmed to include the nearest 4K tokens from the API expression. A maximum generation length is 128 tokens. We report the results of the greedy search.
During the post-processing phase 
we check if the source code following target code completion is being generated; if so, the generation is truncated accordingly.
We transform the newly generated text, into an AST to extract API expressions~\citep{ding2023crosscodeeval}. If no API expressions are identifiable, the generation is left unchanged. We apply the same post-processing on the target completions before calculating the evaluation metrics.

\section{Results}
\label{sec:results}

\paragraph{Library evolution impacts code LLM performance}
We evaluate how the performance of code LLMs changes as libraries evolve by performing code completions for eight libraries: \texttt{torch, torchvision, scipy, pandas, pillow, pyyaml, and tqdm}. This evaluation uses StarCoder2, Mistral, and GPT4o-mini. As shown in the first two columns of Table~\ref{tab:prompting_table}, these models employ different completion strategies: StarCoder2 uses fill-in-the-middle, Mistral utilizes left-context only, while GPT4o-mini follows an instruction-based approach with a one-shot example. All eight libraries are benchmarked in realistic scenarios. Figure~\ref{fig:zero_shot_eval} shows that the developer experience can vary significantly across all models and libraries as public libraries evolve, highlighting the need for better model adaptation to API changes.

\begin{figure*}[ht]
    \centering
    \begin{subfigure}[b]{0.30\textwidth}
        \includegraphics[width=\textwidth]{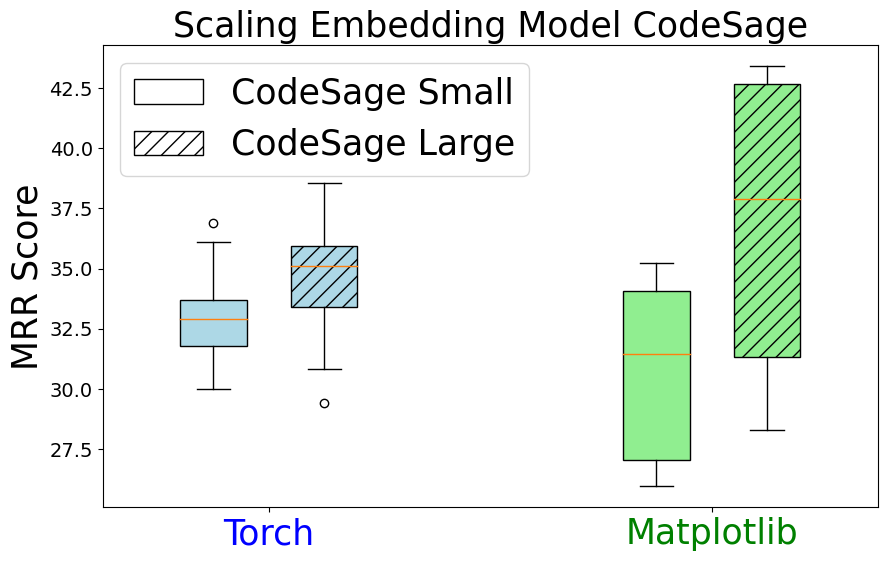}
        \caption{MRR Scores for CodeSage Small and Large Models}
        \label{fig:scaling_codesage}
    \end{subfigure}
    \hfill
    \begin{subfigure}[b]{0.30\textwidth}
        \includegraphics[width=\textwidth]{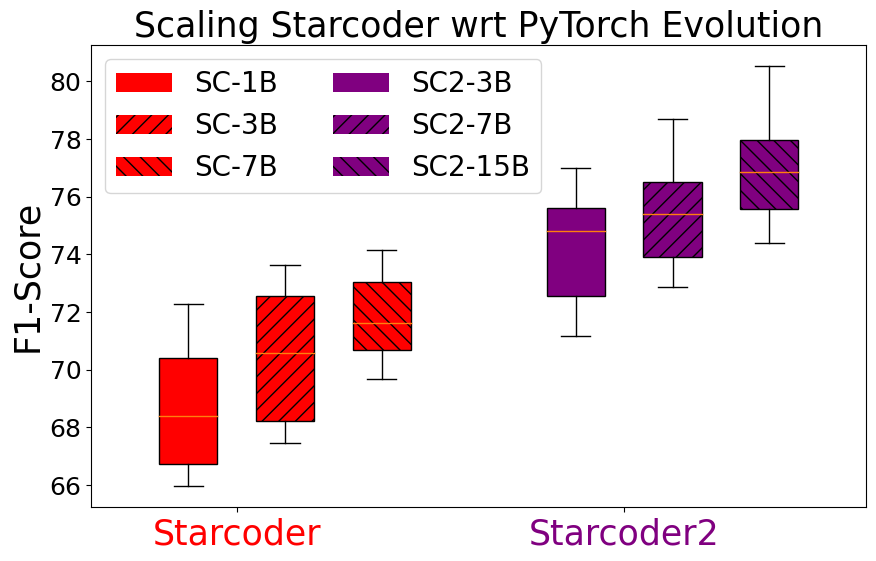}
        \caption{F1 Scores for Starcoder2 and Starcoder Models with respect to PyTorch}
        \label{fig:scaling_starcoder}
    \end{subfigure}
    \hfill
    \begin{subfigure}[b]{0.30\textwidth}
        \includegraphics[width=\textwidth]{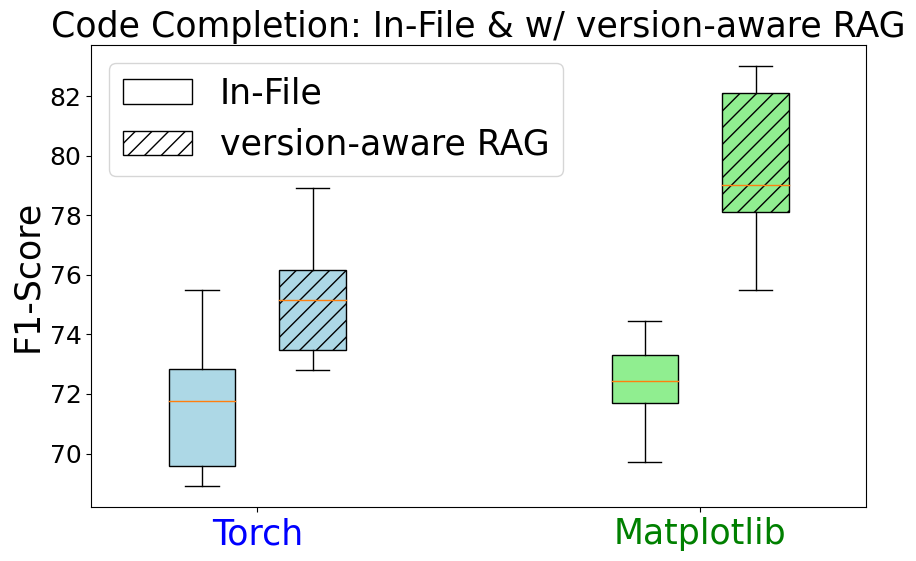}
        \caption{In-File vs Version-Aware RAG Performance for Starcoder2}
        \label{fig:infile_rag}
    \end{subfigure}

    \vspace{10pt} 

    \begin{subfigure}[b]{0.30\textwidth}
        \includegraphics[width=\textwidth]{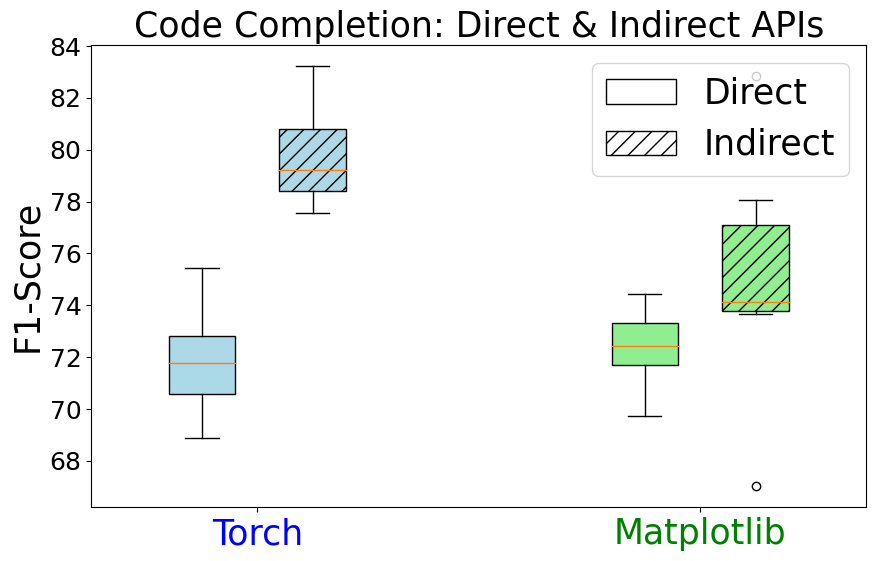}
        \caption{Direct vs Indirect Code Completion for Starcoder2}
        \label{fig:direct_indirect}
    \end{subfigure}
    \hfill
    \begin{subfigure}[b]{0.30\textwidth}
        \includegraphics[width=\textwidth]{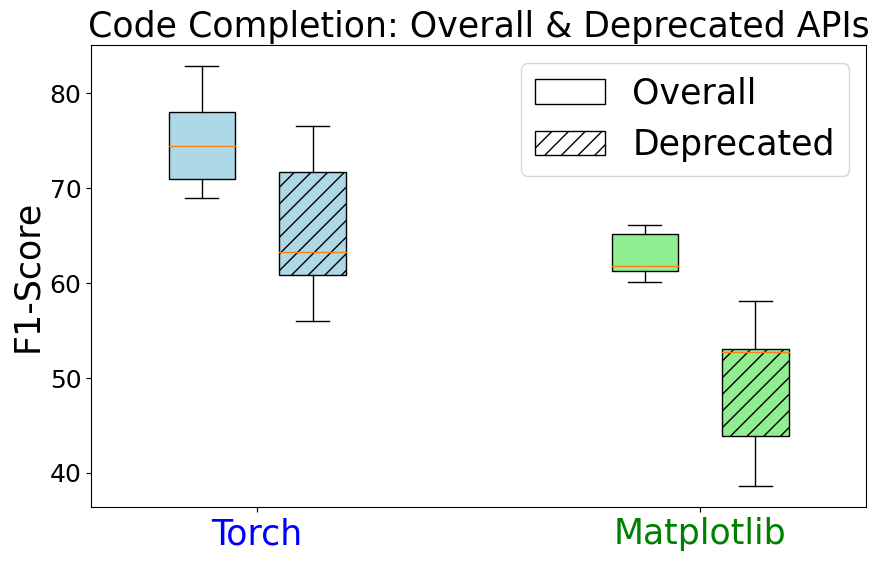}
        \caption{Overall vs Deprecated Set Code Completion for Starcoder2}
        \label{fig:overall_deprecated}
    \end{subfigure}
    \hfill
    \begin{subfigure}[b]{0.30\textwidth}
        \includegraphics[width=\textwidth]{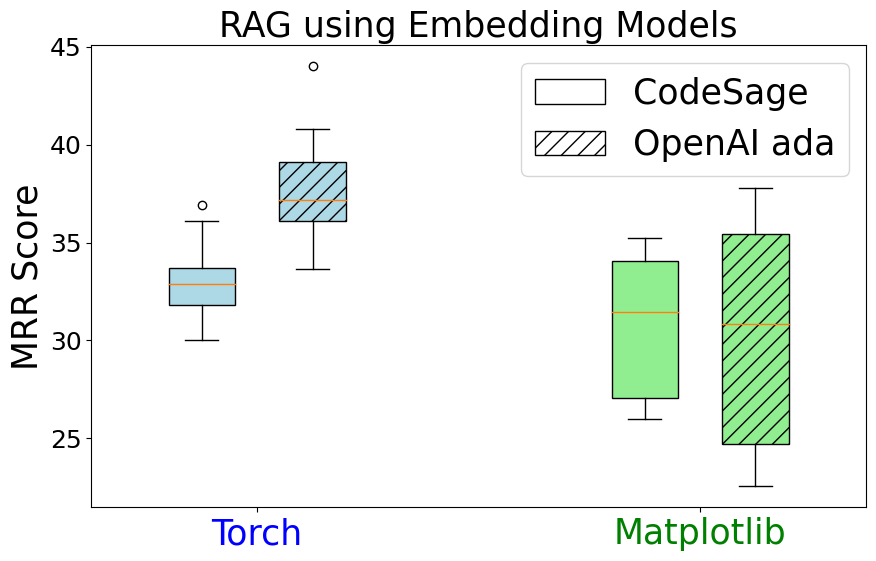}
        \caption{MRR Scores for open CodeSage and Open-Ada-002 models.}
        \label{fig:retrieval}
    \end{subfigure}
    \caption{Detailed analysis of the impact on code completion as \texttt{PyTorch} and \texttt{Matplotlib} evolve using API-completion and documentation retrieval tasks.}
    \label{fig:combined_plots}
    \vspace{-2pt}
\end{figure*}

\paragraph{Version aware contexts enhance code LLM performance}
Table~\ref{tab:prompting_table} demonstrates a clear improvement in model performance as additional contextual information is provided during the code completion task. In the baseline \emph{In-File} setting, where the models rely solely on the code context within a file, the performance is the lowest across all models. Introducing version awareness significantly enhances accuracy, as models can better disambiguate API usage across different library versions. The most notable improvements occur in the \emph{Version-Aware RAG} setting, where documentation relevant to the specific library version is retrieved and used to further refine API completions. This enriched context enables models to generate more precise completions by taking into account the evolving API landscape. These results emphasize the critical role that version-specific and dependency-aware contexts play in improving the accuracy and reliability of code completions.

Furthermore, Figure~\ref{fig:infile_rag} visualizes the impact of using version-aware RAG compared to in-file context settings for the StarCoder2 model, focusing on the evolution of \texttt{PyTorch} and \texttt{Matplotlib}. Although version-aware RAG consistently improves the performance of version-specific code completions, it does not fully address the model's internal bias toward certain versions of the libraries. The box plot highlights variance in the model's predictions, suggesting that despite the enhanced contextual information, underlying biases in the LLM remain, likely stemming from the uneven distribution of training data across library versions.

\paragraph{Library evolution impacts documentation retrieval using embedding models} 
We measure the performance of version-specific documentation retrieval as libraries evolve, using embedding models. As shown in Figure~\ref{fig:retrieval}, we benchmark public CodeSage-Small and the closed-source OpenAI Ada models on \texttt{Torch} and \texttt{Matplotlib}. We observe that the performance of these embedding models fluctuates with the evolution of libraries. This insight sheds light on why version-aware RAG enhances performance for version-specific code completion tasks, but cannot fully resolve the variance in performance across different library versions. The embedding models themselves exhibit bias toward certain library versions, explaining the persistent performance gaps in version-specific code completions.

\paragraph{Impact of scaling and model updates on handling evolving APIs} We evaluate the impact of scaling both embedding models (Figure~\ref{fig:scaling_codesage}) and code-completion models (Figure~\ref{fig:scaling_starcoder}) across different sizes to observe improved retrieval and code completion performance respectively. However, performance still fluctuates as libraries evolve, suggesting that while scaling improves results, it does not fully address the challenges posed by evolving libraries. Furthermore, updating to newer versions, such as from StarCoder-7B to StarCoder2-7B shows that while performance improvements overall are observed (due to better training methods), these do not address the biases introduced by the evolution of public libraries. This points to a need for more specialized training techniques, such as fine-tuning using versioned datasets or incorporating explicit temporal data about API evolution.

\begin{table*}[th]
    \centering
    \begin{minipage}[t]{0.31\textwidth}
        \subcaption{StarCoder2 (Model Release: 2024) on  Matplotlib Deprecated APIs.}
        \vspace{4pt}
        \label{tab:matplotlib_deprecated}
        \resizebox{\textwidth}{!}{
        \begin{tabular}{|c|c|c|}
            \hline
            \textbf{Version} & \textbf{Deprecated} & \textbf{Overall} \\
            \textbf{ Year} & \textbf{API Score} & \textbf{Score} \\
            \hline
            2019 & \underline{38.58} & \textbf{61.83} \\
            2020 & \underline{43.93} & \textbf{60.12} \\
            2021 & 53.01 & \textbf{61.31} \\
            2022 & 52.74 & \textbf{65.15} \\
            2023 & 57.14 & \textbf{66.08} \\
            \hline
        \end{tabular}
        }
    \end{minipage}
    \hfill
    \begin{minipage}[t]{0.31\textwidth}
        \subcaption{CodeGen 1.0 (Knowledge Cutoff: 2022) on  Matplotlib Introduced APIs.}
        \vspace{4pt}
        \label{tab:matplotlib_new}
        \resizebox{\textwidth}{!}{
        \begin{tabular}{|c|c|c|}
            \hline
            \textbf{Version} & \textbf{Introduced} & \textbf{Overall} \\
            \textbf{ Year} & \textbf{API Score} & \textbf{Score} \\
            \hline
            2020 & 53.14 & \textbf{62.89} \\
            2021 & 54.23 & \textbf{62.85} \\
            2022 & 56.29 & \textbf{60.04} \\
            2023 & \underline{44.08} & \textbf{59.44} \\
            2024 & \underline{41.37} & \textbf{58.57} \\
            \hline
        \end{tabular}
        }
    \end{minipage}
    \hfill
    \begin{minipage}[t]{0.34\textwidth}
        \subcaption{StarCoder2 (Knowledge Cutoff: 2024) on PyTorch Introduced/Deprecated APIs.}
        \vspace{4pt}
        \label{tab:pytorch_ndm}
        \resizebox{\textwidth}{!}{
        \begin{tabular}{|c|c|c|}
            \hline
            \textbf{Version} & \textbf{Deprecated/Intro-} & \textbf{Overall} \\
            \textbf{Year} & \textbf{-duced API Score} & \textbf{Score} \\
            \hline
            2018 & 68.78 & \textbf{71.06} \\
            2020 & 59.10 & \textbf{75.45} \\
            2021 & 68.13 & \textbf{72.84} \\
            2022 & 60.93 & \textbf{71.02} \\
            2023 & 67.13 & \textbf{72.82} \\
            \hline
        \end{tabular}
        }
    \end{minipage}
    \vspace{-4pt}
    \caption{Performance comparison of models on different API sets across library versions. Underlined scores indicate a significant performance drop while bold scores are maximum across the two settings considered.
    }
    \vspace{-4pt}
    \label{tab:api_comparison}
\end{table*}

\paragraph{Direct vs indirect API completion}
We evaluate the performance of direct and indirect code completions as libraries evolve in Figure~\ref{fig:direct_indirect}. 
The model exhibits better performance for indirect code completions than direct code completions. 
From subsection~\ref{subsec:api_classification}, we know that every indirect code completion example contains a corresponding parent direct API call in the left context. 
We posit that the code LLM can understand that it needs to perform code completion so that the generated code serves as an attribute related to this version-specific parent direct API call present in the left context.

\paragraph{Impact of different APIs on code completion}
In Figure~\ref{fig:overall_deprecated}, we compare overall performance with the subset of deprecated APIs for code completions in \texttt{PyTorch} and \texttt{Matplotlib}, focusing on a controlled setting. The results show that code LLMs struggle with deprecated APIs, consistently performing worse compared to the overall set. This observation aligns with our qualitative results that models prefer API calls for newer versions of the library (see Figure~\ref{fig:context_example} and~\ref{fig:real_torch_example}). To verify this in a realistic scenario, Table~\ref{tab:pytorch_ndm} provides a version-by-version analysis for PyTorch, comparing newly introduced and deprecated APIs. We observe consistently lower performance on deprecated APIs, except for 2018, a year dominated by newly introduced APIs, which are now widely adopted. These findings demonstrate that rapid changes to API impact models' ability to complete code accurately, with deprecated APIs posing particular challenges.

\paragraph{Temporal analysis of model performance on introduced and deprecated APIs}
The tables compare the performance of StarCoder2 and CodeGen-1.0 on \texttt{Matplotlib}’s deprecated and introduced APIs, respectively, in a controlled setting. Table~\ref{tab:matplotlib_deprecated} shows that StarCoder2 struggles with older, deprecated APIs from 2019 and 2020, indicating that API forgetting contributes to version-specific performance drops. In contrast, Table~\ref{tab:matplotlib_new} highlights a sharp decline in CodeGen-1.0’s performance on introduced APIs from 2023 and 2024, revealing its 2022 knowledge cutoff. This suggests that such tasks could be used to estimate a model's knowledge cutoff. In general, the results underscore the need to develop model training techniques to better handle library evolution.

\section{Related Works}
\label{sec:related_works}

Large language models for code excel in various software development tasks \citep{yan2023codetransocean, lachaux2020unsupervised, roziere2021leveraging, min2023beyond}, facilitating the developments of coding assistants.
Similarly, developments in code embedding models used for retrieval \citep{robertson2009probabilistic, guo2022unixcoder, zhang2024codesage, openai-embedding-ada-002}
have further enhanced LLMs' capabilities. 
In this journey, evaluation benchmarks have played a pivotal role 
with numerous works developing benchmarks to evaluate code LLMs~\citep{humaneval_x, cassano2023multiple, hendrycks2021measuring, lu2021codexglue, puri2021codenet, clement-etal-2021-long,ding2023static,wang2023recode, lu-etal-2022-reacc}. These studies typically assess code completion abilities given local file contexts, both in-file~\citep{chen2021evaluating, athiwaratkun2022multi, lu2021codexglue} and repository-level~\citep{ding2023crosscodeeval, zhang2023repocoder, liu2023repobench,ding2024cocomic}. However, they do not fully encompass the complexities of real-world software development, which requires extensive use of public libraries. Some works have explored code completion involving public libraries~\citep{liao2023context, zan2022cert, qin2023toolllm, patil2023gorilla}, but they do not address the rapidly evolving nature of these libraries. To fill this gap, we introduce \schemelong that evaluates the performance of LLMs on code completion across multiple versions of public libraries, capturing their evolution and reflecting real-world scenarios where developers interact with different versions of the same library.

\section{Conclusion}

In this paper, we introduced \schemelong, a comprehensive benchmark specifically designed to assess the performance of Code Large Language Models (code LLMs) in code completion tasks as public libraries evolve. Our results demonstrate significant variability in LLM performance based on the API version, highlighting the challenges of handling library evolution.
The findings underscore the necessity for future advancements in code completion technologies to consider the dynamic nature of public libraries, aiming to improve developer productivity and accuracy in real-world settings.

\emph{Acknowledgements}: The authors also thank Ming Tan and Hantian Ding for their constructive feedback provided during the paper
writing process. Additionally, we would like to express
gratitude to some other team members from Amazon Q Developer for their insightful discussions,
which have contributed to the refinement of our work.


\bibliography{main.bib}

\begin{thebibliography}{38}
\providecommand{\natexlab}[1]{#1}

\bibitem[{Anthropic(2023)}]{claude}
Anthropic. 2023.
\newblock \href {https://www.anthropic.com/news/claude-2} {\emph{Claude}}.

\bibitem[{Athiwaratkun et~al.(2023)Athiwaratkun, Gouda, Wang, Li, Tian, Tan, Ahmad, Wang, Sun, Shang, Gonugondla, Ding, Kumar, Fulton, Farahani, Jain, Giaquinto, Qian, Ramanathan, Nallapati, Ray, Bhatia, Sengupta, Roth, and Xiang}]{athiwaratkun2022multi}
Ben Athiwaratkun, Sanjay~Krishna Gouda, Zijian Wang, Xiaopeng Li, Yuchen Tian, Ming Tan, Wasi~Uddin Ahmad, Shiqi Wang, Qing Sun, Mingyue Shang, Sujan~Kumar Gonugondla, Hantian Ding, Varun Kumar, Nathan Fulton, Arash Farahani, Siddhartha Jain, Robert Giaquinto, Haifeng Qian, Murali~Krishna Ramanathan, Ramesh Nallapati, Baishakhi Ray, Parminder Bhatia, Sudipta Sengupta, Dan Roth, and Bing Xiang. 2023.
\newblock \href {https://openreview.net/forum?id=Bo7eeXm6An8} {Multi-lingual evaluation of code generation models}.
\newblock In \emph{The Eleventh International Conference on Learning Representations}.

\bibitem[{Bavarian et~al.(2022)Bavarian, Jun, Tezak, Schulman, McLeavey, Tworek, and Chen}]{bavarian2022efficient}
Mohammad Bavarian, Heewoo Jun, Nikolas Tezak, John Schulman, Christine McLeavey, Jerry Tworek, and Mark Chen. 2022.
\newblock \href {https://arxiv.org/abs/2207.14255} {Efficient training of language models to fill in the middle}.
\newblock \emph{arXiv preprint arXiv:2207.14255}.

\bibitem[{Cassano et~al.(2023)Cassano, Gouwar, Nguyen, Nguyen, Phipps-Costin, Pinckney, Yee, Zi, Anderson, Feldman, Guha, Greenberg, and Jangda}]{cassano2023multiple}
Federico Cassano, John Gouwar, Daniel Nguyen, Sydney Nguyen, Luna Phipps-Costin, Donald Pinckney, Ming-Ho Yee, Yangtian Zi, Carolyn~Jane Anderson, Molly~Q Feldman, Arjun Guha, Michael Greenberg, and Abhinav Jangda. 2023.
\newblock \href {https://doi.org/10.1109/TSE.2023.3267446} {Multipl-e: A scalable and polyglot approach to benchmarking neural code generation}.
\newblock \emph{IEEE Transactions on Software Engineering}, 49(7):3675--3691.

\bibitem[{Chen et~al.(2021)Chen, Tworek, Jun, Yuan, Pinto, Kaplan, Edwards, Burda, Joseph, Brockman et~al.}]{chen2021evaluating}
Mark Chen, Jerry Tworek, Heewoo Jun, Qiming Yuan, Henrique Ponde de~Oliveira Pinto, Jared Kaplan, Harri Edwards, Yuri Burda, Nicholas Joseph, Greg Brockman, et~al. 2021.
\newblock \href {https://arxiv.org/abs/2107.03374} {Evaluating large language models trained on code}.
\newblock \emph{ArXiv preprint}, abs/2107.03374.

\bibitem[{Clement et~al.(2021)Clement, Lu, Liu, Tufano, Drain, Duan, Sundaresan, and Svyatkovskiy}]{clement-etal-2021-long}
Colin Clement, Shuai Lu, Xiaoyu Liu, Michele Tufano, Dawn Drain, Nan Duan, Neel Sundaresan, and Alexey Svyatkovskiy. 2021.
\newblock \href {https://doi.org/10.18653/v1/2021.emnlp-main.387} {Long-range modeling of source code files with e{WASH}: Extended window access by syntax hierarchy}.
\newblock In \emph{Proceedings of the 2021 Conference on Empirical Methods in Natural Language Processing}, pages 4713--4722, Online and Punta Cana, Dominican Republic. Association for Computational Linguistics.

\bibitem[{Ding et~al.(2023{\natexlab{a}})Ding, Kumar, Tian, Wang, Kwiatkowski, Li, Ramanathan, Ray, Bhatia, and Sengupta}]{ding2023static}
Hantian Ding, Varun Kumar, Yuchen Tian, Zijian Wang, Rob Kwiatkowski, Xiaopeng Li, Murali~Krishna Ramanathan, Baishakhi Ray, Parminder Bhatia, and Sudipta Sengupta. 2023{\natexlab{a}}.
\newblock \href {https://doi.org/10.18653/v1/2023.acl-industry.34} {A static evaluation of code completion by large language models}.
\newblock In \emph{Proceedings of the 61st Annual Meeting of the Association for Computational Linguistics (Volume 5: Industry Track)}, pages 347--360, Toronto, Canada. Association for Computational Linguistics.

\bibitem[{Ding et~al.(2024)Ding, Wang, Ahmad, Ramanathan, Nallapati, Bhatia, Roth, and Xiang}]{ding2024cocomic}
Yangruibo Ding, Zijian Wang, Wasi~U. Ahmad, Murali~Krishna Ramanathan, Ramesh Nallapati, Parminder Bhatia, Dan Roth, and Bing Xiang. 2024.
\newblock \href {https://aclanthology.org/2024.lrec-main.305} {{C}o{C}o{MIC}: Code completion by jointly modeling in-file and cross-file context}.
\newblock In \emph{Proceedings of the 2024 Joint International Conference on Computational Linguistics, Language Resources and Evaluation (LREC-COLING 2024)}, pages 3433--3445, Torino, Italia. ELRA and ICCL.

\bibitem[{Ding et~al.(2023{\natexlab{b}})Ding, Wang, Ahmad, Ding, Tan, Jain, Ramanathan, Nallapati, Bhatia, Roth, and Xiang}]{ding2023crosscodeeval}
Yangruibo Ding, Zijian Wang, Wasi~Uddin Ahmad, Hantian Ding, Ming Tan, Nihal Jain, Murali~Krishna Ramanathan, Ramesh Nallapati, Parminder Bhatia, Dan Roth, and Bing Xiang. 2023{\natexlab{b}}.
\newblock \href {https://openreview.net/forum?id=wgDcbBMSfh} {Crosscodeeval: A diverse and multilingual benchmark for cross-file code completion}.
\newblock In \emph{Thirty-seventh Conference on Neural Information Processing Systems Datasets and Benchmarks Track}.

\bibitem[{Guo et~al.(2022)Guo, Lu, Duan, Wang, Zhou, and Yin}]{guo2022unixcoder}
Daya Guo, Shuai Lu, Nan Duan, Yanlin Wang, Ming Zhou, and Jian Yin. 2022.
\newblock Unixcoder: Unified cross-modal pre-training for code representation.
\newblock In \emph{Proceedings of the 60th Annual Meeting of the Association for Computational Linguistics (Volume 1: Long Papers)}, pages 7212--7225.

\bibitem[{Hendrycks et~al.(2021)Hendrycks, Basart, Kadavath, Mazeika, Arora, Guo, Burns, Puranik, He, Song, and Steinhardt}]{hendrycks2021measuring}
Dan Hendrycks, Steven Basart, Saurav Kadavath, Mantas Mazeika, Akul Arora, Ethan Guo, Collin Burns, Samir Puranik, Horace He, Dawn Song, and Jacob Steinhardt. 2021.
\newblock \href {https://openreview.net/forum?id=sD93GOzH3i5} {Measuring coding challenge competence with {APPS}}.
\newblock In \emph{Thirty-fifth Conference on Neural Information Processing Systems Datasets and Benchmarks Track (Round 2)}.

\bibitem[{Jiang et~al.(2023)Jiang, Sablayrolles, Mensch, Bamford, Chaplot, Casas, Bressand, Lengyel, Lample, Saulnier et~al.}]{jiang2023mistral}
Albert~Q Jiang, Alexandre Sablayrolles, Arthur Mensch, Chris Bamford, Devendra~Singh Chaplot, Diego de~las Casas, Florian Bressand, Gianna Lengyel, Guillaume Lample, Lucile Saulnier, et~al. 2023.
\newblock Mistral 7b.
\newblock \emph{arXiv preprint arXiv:2310.06825}.

\bibitem[{Jimenez et~al.(2024)Jimenez, Yang, Wettig, Yao, Pei, Press, and Narasimhan}]{jimenez2024swebench}
Carlos~E Jimenez, John Yang, Alexander Wettig, Shunyu Yao, Kexin Pei, Ofir Press, and Karthik~R Narasimhan. 2024.
\newblock \href {https://openreview.net/forum?id=VTF8yNQM66} {{SWE}-bench: Can language models resolve real-world github issues?}
\newblock In \emph{The Twelfth International Conference on Learning Representations}.

\bibitem[{Li et~al.(2023)Li, Allal, Zi, Muennighoff, Kocetkov, Mou, Marone, Akiki, Li, Chim et~al.}]{li2023starcoder}
Raymond Li, Loubna~Ben Allal, Yangtian Zi, Niklas Muennighoff, Denis Kocetkov, Chenghao Mou, Marc Marone, Christopher Akiki, Jia Li, Jenny Chim, et~al. 2023.
\newblock \href {https://arxiv.org/abs/2305.06161} {Starcoder: may the source be with you!}
\newblock \emph{arXiv preprint arXiv:2305.06161}.

\bibitem[{Li et~al.(2022)Li, Choi, Chung, Kushman, Schrittwieser, Leblond, Eccles, Keeling, Gimeno, Lago, Hubert, Choy, de~Masson~d’Autume, Babuschkin, Chen, Huang, Welbl, Gowal, Cherepanov, Molloy, Mankowitz, Robson, Kohli, de~Freitas, Kavukcuoglu, and Vinyals}]{doi:10.1126/science.abq1158}
Yujia Li, David Choi, Junyoung Chung, Nate Kushman, Julian Schrittwieser, R{\'e}mi Leblond, Tom Eccles, James Keeling, Felix Gimeno, Agustin~Dal Lago, Thomas Hubert, Peter Choy, Cyprien de~Masson~d’Autume, Igor Babuschkin, Xinyun Chen, Po-Sen Huang, Johannes Welbl, Sven Gowal, Alexey Cherepanov, James Molloy, Daniel~J. Mankowitz, Esme~Sutherland Robson, Pushmeet Kohli, Nando de~Freitas, Koray Kavukcuoglu, and Oriol Vinyals. 2022.
\newblock \href {https://doi.org/10.1126/science.abq1158} {Competition-level code generation with alphacode}.
\newblock \emph{Science}, 378(6624):1092--1097.

\bibitem[{Liao et~al.(2023)Liao, Pan, Huang, Ren, Xing, Jin, and Li}]{liao2023context}
Dianshu Liao, Shidong Pan, Qing Huang, Xiaoxue Ren, Zhenchang Xing, Huan Jin, and Qinying Li. 2023.
\newblock Context-aware code generation framework for code repositories: Local, global, and third-party library awareness.
\newblock \emph{arXiv preprint arXiv:2312.05772}.

\bibitem[{Liu et~al.(2024)Liu, Xu, and McAuley}]{liu2023repobench}
Tianyang Liu, Canwen Xu, and Julian McAuley. 2024.
\newblock \href {https://arxiv.org/abs/2306.03091} {Repobench: Benchmarking repository-level code auto-completion systems}.
\newblock In \emph{International Conference on Learning Representations}.

\bibitem[{Lozhkov et~al.(2024)Lozhkov, Li, Allal, Cassano, Lamy-Poirier, Tazi, Tang, Pykhtar, Liu, Wei, Liu, Tian, Kocetkov, Zucker, Belkada, Wang, Liu, Abulkhanov, Paul, Li, Li, Risdal, Li, Zhu, Zhuo, Zheltonozhskii, Dade, Yu, Krauß, Jain, Su, He, Dey, Abati, Chai, Muennighoff, Tang, Oblokulov, Akiki, Marone, Mou, Mishra, Gu, Hui, Dao, Zebaze, Dehaene, Patry, Xu, McAuley, Hu, Scholak, Paquet, Robinson, Anderson, Chapados, Patwary, Tajbakhsh, Jernite, Ferrandis, Zhang, Hughes, Wolf, Guha, von Werra, and de~Vries}]{lozhkov2024starcoder}
Anton Lozhkov, Raymond Li, Loubna~Ben Allal, Federico Cassano, Joel Lamy-Poirier, Nouamane Tazi, Ao~Tang, Dmytro Pykhtar, Jiawei Liu, Yuxiang Wei, Tianyang Liu, Max Tian, Denis Kocetkov, Arthur Zucker, Younes Belkada, Zijian Wang, Qian Liu, Dmitry Abulkhanov, Indraneil Paul, Zhuang Li, Wen-Ding Li, Megan Risdal, Jia Li, Jian Zhu, Terry~Yue Zhuo, Evgenii Zheltonozhskii, Nii Osae~Osae Dade, Wenhao Yu, Lucas Krauß, Naman Jain, Yixuan Su, Xuanli He, Manan Dey, Edoardo Abati, Yekun Chai, Niklas Muennighoff, Xiangru Tang, Muhtasham Oblokulov, Christopher Akiki, Marc Marone, Chenghao Mou, Mayank Mishra, Alex Gu, Binyuan Hui, Tri Dao, Armel Zebaze, Olivier Dehaene, Nicolas Patry, Canwen Xu, Julian McAuley, Han Hu, Torsten Scholak, Sebastien Paquet, Jennifer Robinson, Carolyn~Jane Anderson, Nicolas Chapados, Mostofa Patwary, Nima Tajbakhsh, Yacine Jernite, Carlos~Muñoz Ferrandis, Lingming Zhang, Sean Hughes, Thomas Wolf, Arjun Guha, Leandro von Werra, and Harm de~Vries. 2024.
\newblock \href {https://arxiv.org/abs/2402.19173} {Starcoder 2 and the stack v2: The next generation}.
\newblock \emph{Preprint}, arXiv:2402.19173.

\bibitem[{Lu et~al.(2022)Lu, Duan, Han, Guo, Hwang, and Svyatkovskiy}]{lu-etal-2022-reacc}
Shuai Lu, Nan Duan, Hojae Han, Daya Guo, Seung-won Hwang, and Alexey Svyatkovskiy. 2022.
\newblock \href {https://doi.org/10.18653/v1/2022.acl-long.431} {{R}e{ACC}: A retrieval-augmented code completion framework}.
\newblock In \emph{Proceedings of the 60th Annual Meeting of the Association for Computational Linguistics (Volume 1: Long Papers)}, pages 6227--6240, Dublin, Ireland. Association for Computational Linguistics.

\bibitem[{Lu et~al.(2021)Lu, Guo, Ren, Huang, Svyatkovskiy, Blanco, Clement, Drain, Jiang, Tang, Li, Zhou, Shou, Zhou, Tufano, GONG, Zhou, Duan, Sundaresan, Deng, Fu, and LIU}]{lu2021codexglue}
Shuai Lu, Daya Guo, Shuo Ren, Junjie Huang, Alexey Svyatkovskiy, Ambrosio Blanco, Colin Clement, Dawn Drain, Daxin Jiang, Duyu Tang, Ge~Li, Lidong Zhou, Linjun Shou, Long Zhou, Michele Tufano, MING GONG, Ming Zhou, Nan Duan, Neel Sundaresan, Shao~Kun Deng, Shengyu Fu, and Shujie LIU. 2021.
\newblock \href {https://openreview.net/forum?id=6lE4dQXaUcb} {Code{XGLUE}: A machine learning benchmark dataset for code understanding and generation}.
\newblock In \emph{Thirty-fifth Conference on Neural Information Processing Systems Datasets and Benchmarks Track (Round 1)}.

\bibitem[{McDonnell et~al.(2013)McDonnell, Ray, and Kim}]{mcdonnell2013empirical}
Tyler McDonnell, Baishakhi Ray, and Miryung Kim. 2013.
\newblock An empirical study of api stability and adoption in the android ecosystem.
\newblock In \emph{2013 IEEE International Conference on Software Maintenance}, pages 70--79. IEEE.

\bibitem[{Min et~al.(2024)Min, Ding, Buratti, Pujar, Kaiser, Jana, and Ray}]{min2023beyond}
Marcus~J. Min, Yangruibo Ding, Luca Buratti, Saurabh Pujar, Gail Kaiser, Suman Jana, and Baishakhi Ray. 2024.
\newblock \href {https://openreview.net/forum?id=caW7LdAALh} {Beyond accuracy: Evaluating self-consistency of code large language models with identitychain}.
\newblock In \emph{The Twelfth International Conference on Learning Representations}.

\bibitem[{Nijkamp et~al.(2023)Nijkamp, Pang, Hayashi, Tu, Wang, Zhou, Savarese, and Xiong}]{Nijkamp2022ACP}
Erik Nijkamp, Bo~Pang, Hiroaki Hayashi, Lifu Tu, Huan Wang, Yingbo Zhou, Silvio Savarese, and Caiming Xiong. 2023.
\newblock \href {https://openreview.net/forum?id=iaYcJKpY2B_} {Codegen: An open large language model for code with multi-turn program synthesis}.
\newblock In \emph{The Eleventh International Conference on Learning Representations}.

\bibitem[{OpenAI(2022)}]{openai-embedding-ada-002}
OpenAI. 2022.
\newblock Embedding ada-002.
\newblock \url{https://platform.openai.com/docs/guides/embeddings/using-embeddings}.

\bibitem[{OpenAI(2024)}]{openai2024gpt4o}
OpenAI. 2024.
\newblock Hello gpt-4o.
\newblock \url{https://openai.com/index/hello-gpt-4o/}.
\newblock [Accessed: October 10, 2024].

\bibitem[{Patil et~al.(2023)Patil, Zhang, Wang, and Gonzalez}]{patil2023gorilla}
Shishir~G Patil, Tianjun Zhang, Xin Wang, and Joseph~E Gonzalez. 2023.
\newblock Gorilla: Large language model connected with massive apis.
\newblock \emph{arXiv preprint arXiv:2305.15334}.

\bibitem[{Puri et~al.(2021)Puri, Kung, Janssen, Zhang, Domeniconi, Zolotov, Dolby, Chen, Choudhury, Decker, Thost, Buratti, Pujar, Ramji, Finkler, Malaika, and Reiss}]{puri2021codenet}
Ruchir Puri, David~S Kung, Geert Janssen, Wei Zhang, Giacomo Domeniconi, Vladimir Zolotov, Julian Dolby, Jie Chen, Mihir Choudhury, Lindsey Decker, Veronika Thost, Luca Buratti, Saurabh Pujar, Shyam Ramji, Ulrich Finkler, Susan Malaika, and Frederick Reiss. 2021.
\newblock \href {https://openreview.net/forum?id=6vZVBkCDrHT} {Codenet: A large-scale {AI} for code dataset for learning a diversity of coding tasks}.
\newblock In \emph{Thirty-fifth Conference on Neural Information Processing Systems Datasets and Benchmarks Track (Round 2)}.

\bibitem[{Qin et~al.(2024)Qin, Liang, Ye, Zhu, Yan, Lu, Lin, Cong, Tang, Qian, Zhao, Hong, Tian, Xie, Zhou, Gerstein, dahai li, Liu, and Sun}]{qin2023toolllm}
Yujia Qin, Shihao Liang, Yining Ye, Kunlun Zhu, Lan Yan, Yaxi Lu, Yankai Lin, Xin Cong, Xiangru Tang, Bill Qian, Sihan Zhao, Lauren Hong, Runchu Tian, Ruobing Xie, Jie Zhou, Mark Gerstein, dahai li, Zhiyuan Liu, and Maosong Sun. 2024.
\newblock \href {https://openreview.net/forum?id=dHng2O0Jjr} {Tool{LLM}: Facilitating large language models to master 16000+ real-world {API}s}.
\newblock In \emph{The Twelfth International Conference on Learning Representations}.

\bibitem[{Robertson et~al.(2009)Robertson, Zaragoza et~al.}]{robertson2009probabilistic}
Stephen Robertson, Hugo Zaragoza, et~al. 2009.
\newblock The probabilistic relevance framework: Bm25 and beyond.
\newblock \emph{Foundations and Trends{\textregistered} in Information Retrieval}, 3(4):333--389.

\bibitem[{Roziere et~al.(2023)Roziere, Gehring, Gloeckle, Sootla, Gat, Tan, Adi, Liu, Remez, Rapin et~al.}]{roziere2023code}
Baptiste Roziere, Jonas Gehring, Fabian Gloeckle, Sten Sootla, Itai Gat, Xiaoqing~Ellen Tan, Yossi Adi, Jingyu Liu, Tal Remez, J{\'e}r{\'e}my Rapin, et~al. 2023.
\newblock Code llama: Open foundation models for code.
\newblock \emph{arXiv preprint arXiv:2308.12950}.

\bibitem[{Roziere et~al.(2020)Roziere, Lachaux, Chanussot, and Lample}]{lachaux2020unsupervised}
Baptiste Roziere, Marie-Anne Lachaux, Lowik Chanussot, and Guillaume Lample. 2020.
\newblock Unsupervised translation of programming languages.
\newblock In \emph{Proceedings of the 34th International Conference on Neural Information Processing Systems}, NIPS '20, Red Hook, NY, USA. Curran Associates Inc.

\bibitem[{Roziere et~al.(2022)Roziere, Zhang, Charton, Harman, Synnaeve, and Lample}]{roziere2021leveraging}
Baptiste Roziere, Jie Zhang, Francois Charton, Mark Harman, Gabriel Synnaeve, and Guillaume Lample. 2022.
\newblock \href {https://openreview.net/forum?id=cmt-6KtR4c4} {Leveraging automated unit tests for unsupervised code translation}.
\newblock In \emph{International Conference on Learning Representations}.

\bibitem[{Wang et~al.(2023)Wang, Li, Qian, Yang, Wang, Shang, Kumar, Tan, Ray, Bhatia, Nallapati, Ramanathan, Roth, and Xiang}]{wang2023recode}
Shiqi Wang, Zheng Li, Haifeng Qian, Chenghao Yang, Zijian Wang, Mingyue Shang, Varun Kumar, Samson Tan, Baishakhi Ray, Parminder Bhatia, Ramesh Nallapati, Murali~Krishna Ramanathan, Dan Roth, and Bing Xiang. 2023.
\newblock \href {https://doi.org/10.18653/v1/2023.acl-long.773} {{R}e{C}ode: Robustness evaluation of code generation models}.
\newblock In \emph{Proceedings of the 61st Annual Meeting of the Association for Computational Linguistics (Volume 1: Long Papers)}, pages 13818--13843, Toronto, Canada. Association for Computational Linguistics.

\bibitem[{Yan et~al.(2023)Yan, Tian, Li, Chen, and Wang}]{yan2023codetransocean}
Weixiang Yan, Yuchen Tian, Yunzhe Li, Qian Chen, and Wen Wang. 2023.
\newblock \href {https://doi.org/10.18653/v1/2023.findings-emnlp.337} {{C}ode{T}rans{O}cean: A comprehensive multilingual benchmark for code translation}.
\newblock In \emph{Findings of the Association for Computational Linguistics: EMNLP 2023}, pages 5067--5089, Singapore. Association for Computational Linguistics.

\bibitem[{Zan et~al.(2022)Zan, Chen, Yang, Lin, Kim, Guan, Wang, Chen, and Lou}]{zan2022cert}
Daoguang Zan, Bei Chen, Dejian Yang, Zeqi Lin, Minsu Kim, Bei Guan, Yongji Wang, Weizhu Chen, and Jian-Guang Lou. 2022.
\newblock \href {https://doi.org/10.24963/ijcai.2022/329} {Cert: Continual pre-training on sketches for library-oriented code generation}.
\newblock In \emph{Proceedings of the Thirty-First International Joint Conference on Artificial Intelligence, {IJCAI-22}}, pages 2369--2375. International Joint Conferences on Artificial Intelligence Organization.
\newblock Main Track.

\bibitem[{Zhang et~al.(2024)Zhang, Ahmad, Tan, Ding, Nallapati, Roth, Ma, and Xiang}]{zhang2024codesage}
Dejiao Zhang, Wasi Ahmad, Ming Tan, Hantian Ding, Ramesh Nallapati, Dan Roth, Xiaofei Ma, and Bing Xiang. 2024.
\newblock \href {https://openreview.net/forum?id=vfzRRjumpX} {Codesage: Code representation learning at scale}.
\newblock In \emph{The Twelfth International Conference on Learning Representations}.

\bibitem[{Zhang et~al.(2023)Zhang, Chen, Zhang, Keung, Liu, Zan, Mao, Lou, and Chen}]{zhang2023repocoder}
Fengji Zhang, Bei Chen, Yue Zhang, Jacky Keung, Jin Liu, Daoguang Zan, Yi~Mao, Jian-Guang Lou, and Weizhu Chen. 2023.
\newblock \href {https://doi.org/10.18653/v1/2023.emnlp-main.151} {{R}epo{C}oder: Repository-level code completion through iterative retrieval and generation}.
\newblock In \emph{Proceedings of the 2023 Conference on Empirical Methods in Natural Language Processing}, pages 2471--2484, Singapore. Association for Computational Linguistics.

\bibitem[{Zheng et~al.(2023)Zheng, Xia, Zou, Dong, Wang, Xue, Wang, Shen, Wang, Li, Su, Yang, and Tang}]{humaneval_x}
Qinkai Zheng, Xiao Xia, Xu~Zou, Yuxiao Dong, Shan Wang, Yufei Xue, Zihan Wang, Lei Shen, Andi Wang, Yang Li, Teng Su, Zhilin Yang, and Jie Tang. 2023.
\newblock Codegeex: A pre-trained model for code generation with multilingual benchmarking on humaneval-x.
\newblock In \emph{Proceedings of the 29th ACM SIGKDD Conference on Knowledge Discovery and Data Mining}, pages 5673--5684.

\end{thebibliography}

\clearpage
\appendix

\twocolumn[{%
 \centering
 \Large\bf Supplementary Material: Appendices \\ [20pt]
}]

\section{Limitations}\label{sec:limitations}

This study involves a zero-shot approach to evaluate the impact of the evolution of public libraries on Code LLMs. Pre-training a model exclusively with version-specific data from public libraries might help to reduce the version-dependent discrepancies observed in zero-shot settings. Additionally, it is important to acknowledge that CodeLMs are trained on vast repositories of unlabeled code, raising the possibility that the model might have previously encountered some of the evaluation data. This potential overlap should be carefully considered when interpreting the results of this study.

\begin{figure}[h]
    \centering
    \begin{subfigure}[b]{0.48\textwidth}
        \centering
        \includegraphics[width=\textwidth]{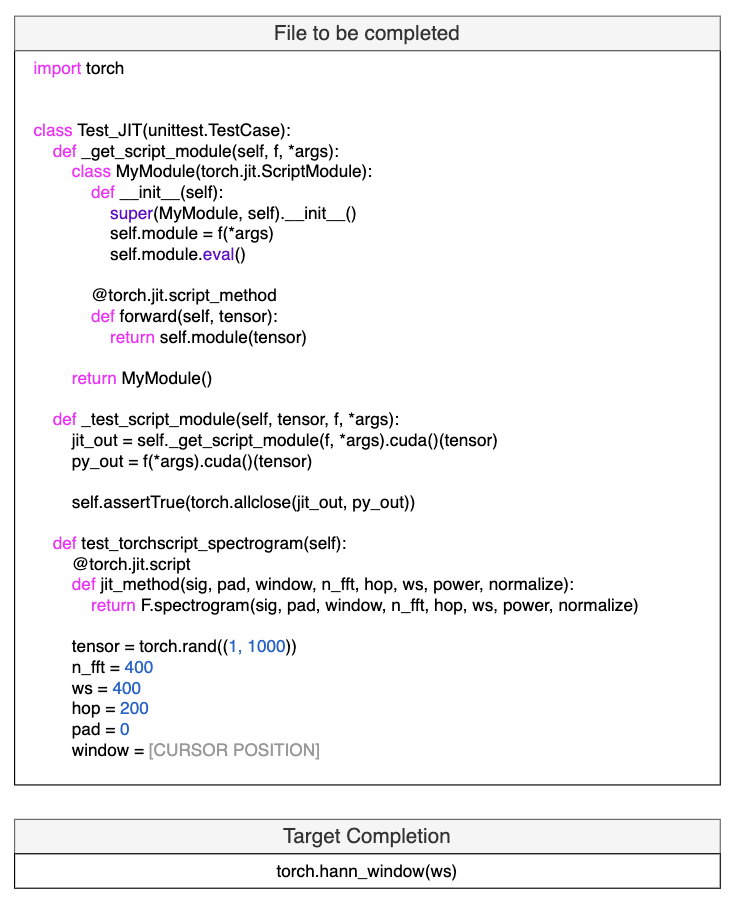}
    \end{subfigure}\\[1ex] 
    \caption{Examples of API evaluation.}
    \label{fig:torch_code_example}
\end{figure}

\begin{figure}[h]
    \centering
\includegraphics[width=0.48\textwidth]{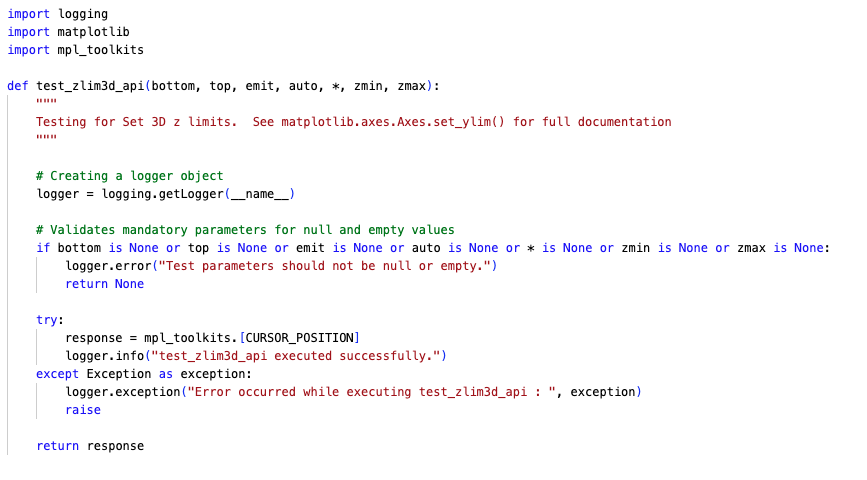}
    \caption{Matplotlib API evaluation example created synthetically from documentation.}
    \label{fig:aws_api_template_example}
\end{figure}

\begin{figure}
    \centering
    \includegraphics[width=0.45\textwidth]{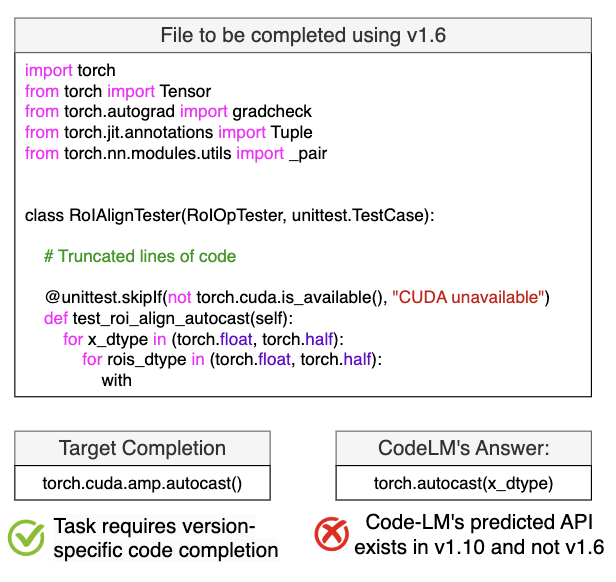}
    \caption{Qualitative example of an error in code completion on \schemelong focused set.}
    \label{fig:real_torch_example}
\end{figure}

\section{\schemelong Generation}\label{sec:data_collection}

\paragraph{Realistic Scenario} Python files using the library are processed to extract library usage patterns by parsing the files into Abstract Syntax Trees (ASTs) using the \texttt{tree-sitter} library. This comprehensive approach allows us to identify syntactic elements such as function calls and import statements specific to the library. The AST is systematically traversed to detect both direct and indirect API calls to the library and its submodules. Direct API calls are identified by their explicit invocation in the source code, typically involving function calls directly on the library modules (e.g., \texttt{torch.nn.Linear}). Indirect API calls are recognized through variables or objects that are assigned to library functions or classes and used later in the code, which requires tracking variable scopes and aliases across the codebase. The broader structural context for a direct API call is determined by locating the closest enclosing syntactic structure, such as a function or a class method, in the AST. This enclosing structure is regarded as the scope of the API call. The entire block of code constituting this scope is extracted as a context. This context includes parameter lists, internal variable declarations, and other code elements within the same block, providing a comprehensive view of how the API is integrated into the function. The context for indirect API calls includes not only the block where the variable is used but also potentially broader code segments that influence or are influenced by the variable use. This methodical extraction of context ensures that each API call, whether direct or indirect, is analyzed within its operational environment concerning left context. 

\paragraph{Controlled Scenario} The library usage data for controlled ablations was created synthetically using the documentation for each version. Each documentation is converted into a code completion example to be used for evaluation using a template (see Figure~\ref{fig:aws_api_template_example}). The template highlights the service name, API description, and mandatory arguments and does not leak the name of the API.

\section{Generating Natural Language Instructions Using Claude}\label{sec:claude_prompt}

\begin{figure}[h]
    \centering
    \includegraphics[width=0.45\textwidth]{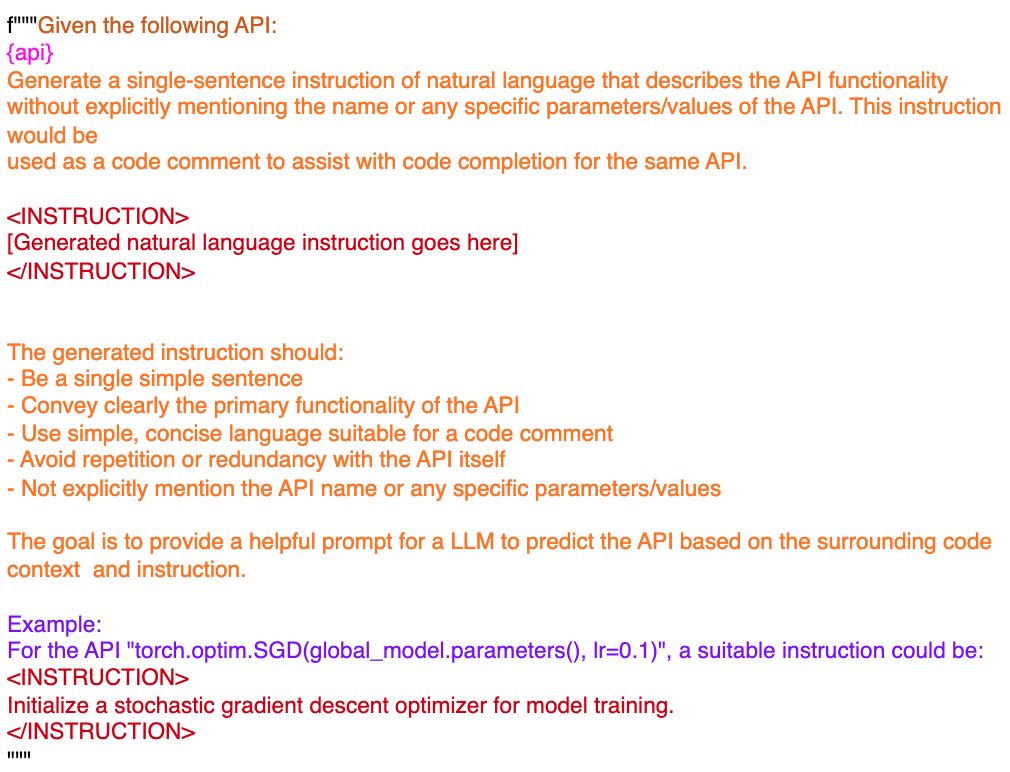}
    \caption{The template used to prompt Claude to create natural language instructions from a target code completion expression. The template explicitly guides the LLM not to provide the name or the arguments required for the code completion in the natural language instruction.}
    \label{fig:claude_prompt}
\end{figure}

\section{Retrieval Performance vs Model Size on \schemelong}\label{sec:scaling_mrr}

\begin{figure}[H]
    \centering
    \includegraphics[width=0.48\textwidth]{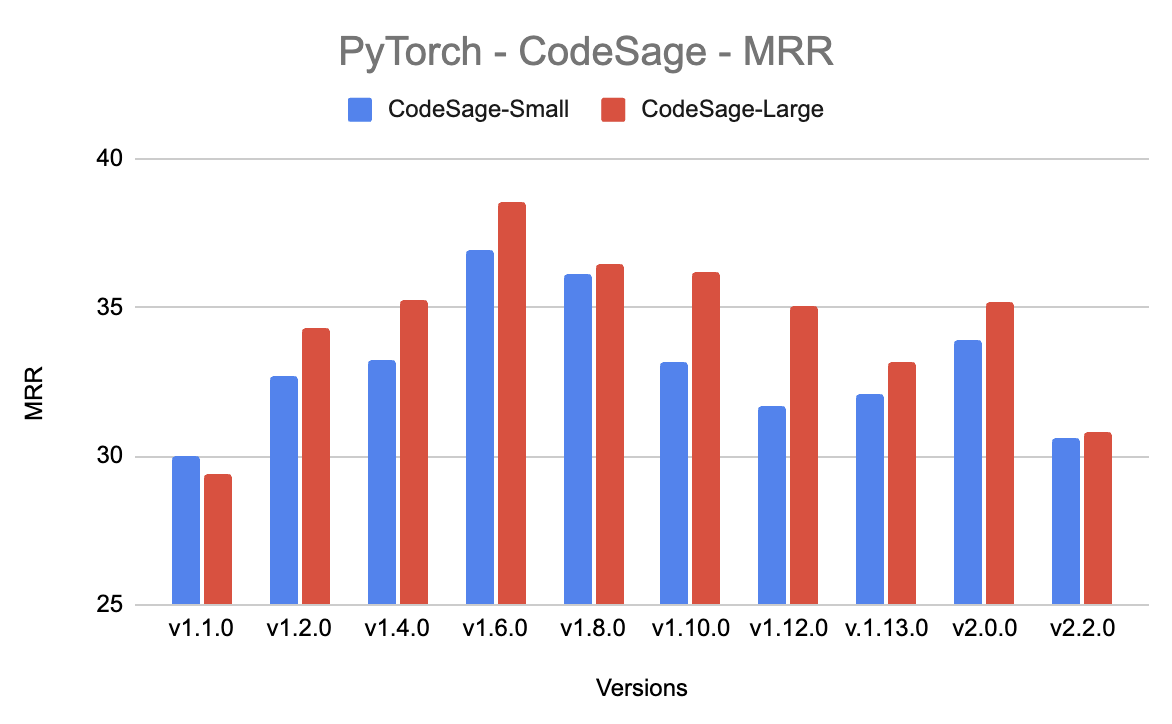}
    \caption{Larger CodeSage models perform better at documentation retrieval}
    \label{fig:enter-label}
\end{figure}

\section{Code Completions Performance vs Model Size on \schemelong}\label{sec:scaling}

\begin{figure}[H]
    \centering
    \begin{subfigure}[b]{0.48\textwidth}
        \centering
        \includegraphics[width=\textwidth]{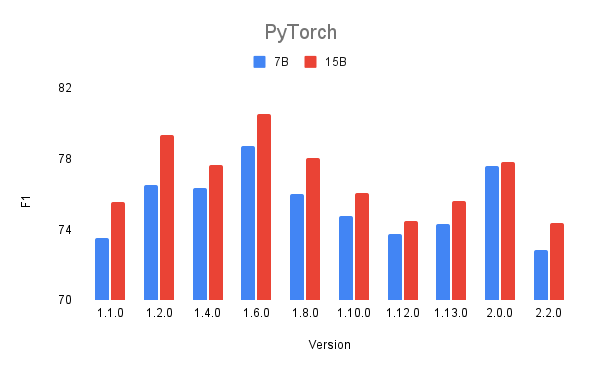}
        \caption{Starcoder2 15B and 7B on PyTorch.}
        \label{fig:left}
    \end{subfigure}
    \hfill
    \begin{subfigure}[b]{0.48\textwidth}
        \centering
        \includegraphics[width=\textwidth]{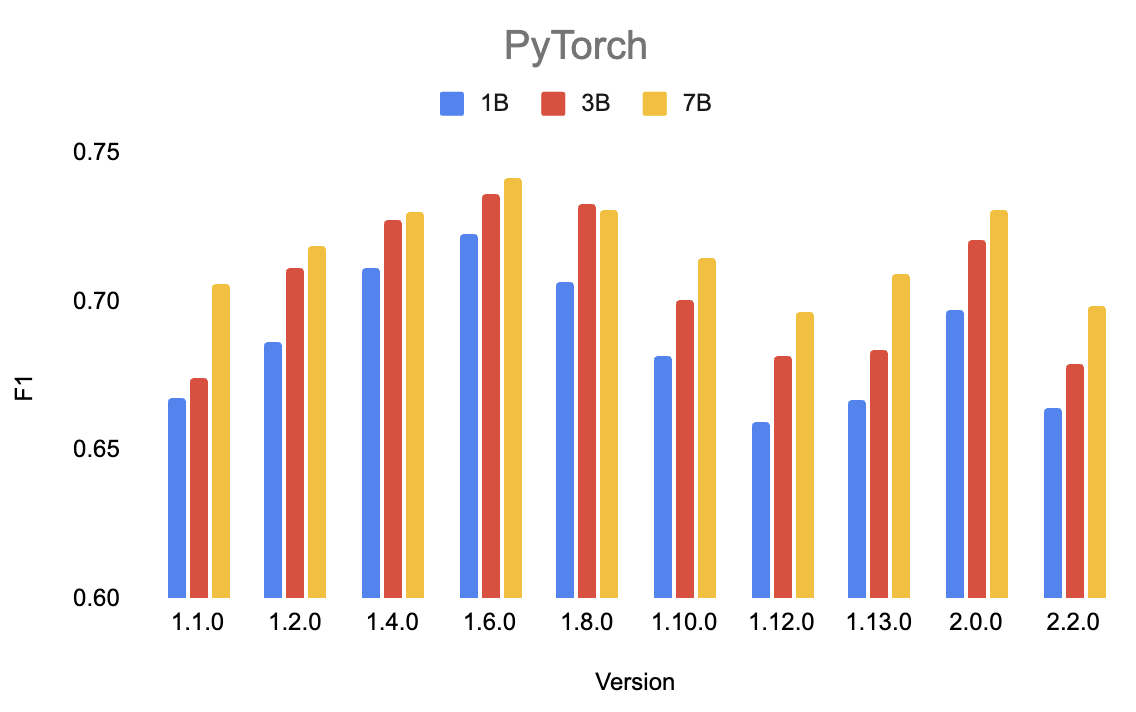}
        \caption{Starcoder 1B, 3B, and 7B on PyTorch.}
        \label{fig:left}
    \end{subfigure}
    \caption{Larger Starcoder and Starcoder2 models perform better at code completions.}
    \label{fig:below_right}
\end{figure}

\section{Detailed Statistics of Data}\label{sec:version_level_stats}

\subsection*{Main Result}
\begin{table}[h!]
\centering
\resizebox{\columnwidth}{!}{
\begin{tabular}{|c|c|c|c|c|c|}
\hline
\textbf{Library} & \textbf{2019} & \textbf{2020} & \textbf{2021} & \textbf{2022} & \textbf{2023} \\ \hline
torch        & 286 & 513 & 708 & 835 & 738 \\ \hline
torchvision  & 16  & 16  & 14  & 16  & 16  \\ \hline
scipy        & 70  & 70  & 70  & 70  & 68  \\ \hline
pandas       & 60  & 60  & 60  & 59  & 60  \\ \hline
matplotlib   & 90  & 90  & 90  & 89  & 90  \\ \hline
pillow       & 40  & 40  & 40  & 40  & 40  \\ \hline
pyyaml       & 8   & 8   & 8   & Na  & 8   \\ \hline
tqdm         & 12  & 12  & 12  & 12  & 12  \\ \hline
\end{tabular}}
\caption{API Examples per Year for Libraries}
\label{tab:additional_libraries_data}
\end{table}

\subsection*{Matplotlib Library Ablation Data}

\begin{table}[H]
\centering

\resizebox{\columnwidth}{!}{
\begin{tabular}{|c|c|c|c|c|c|}
\hline
\textbf{Version} & \textbf{Deprecated} & \textbf{Overall} & \textbf{Direct} & \textbf{Indirect} & \textbf{Introduced} \\ \hline
3\_0\_3 & 40  & 1000 & 297 & 359 & N/A \\ \hline
3\_2\_0 & 40  & 1000 & 154 & 175 & 200 \\ \hline
3\_3\_4 & 40  & 1000 & 339 & 395 & 200 \\ \hline
3\_5\_2 & 40  & 1000 & 291 & 298 & 200 \\ \hline
3\_6\_3 & 40  & 1000 & 84  & 31  & 200 \\ \hline
3\_8\_3 & N/A & 1000 & 37  & 8   & 200 \\ \hline
\end{tabular}}
\caption{Comparison between Matplotlib Focussed vs. Comprehensive dataset.}
\label{tab:matplotlib_data}
\end{table}

\subsection*{PyTorch Library Data}

\begin{table}[H]
\centering
\caption{PyTorch API Usage Data}
\resizebox{\columnwidth}{!}{
\begin{tabular}{|c|c|c|c|c|c|}
\hline
\textbf{Version} & \textbf{Direct} & \textbf{Indirect} & \textbf{Intr/Depr} & \textbf{Deprecated} & \textbf{Overall} \\ \hline
v\_1\_1\_0 & 286 & 174 & 50 & N/A & 1000 \\ \hline
v\_1\_2\_0 & 341 & 189 & 35 & 9 & 1000 \\ \hline
v\_1\_4\_0 & 452 & 262 & 45 & 14 & 1000 \\ \hline
v\_1\_6\_0 & 513 & 300 & 50 & N/A & 1000 \\ \hline
v\_1\_8\_0 & 564 & 353 & 50 & 40 & 1000 \\ \hline
v\_1\_10\_0 & 708 & 466 & 50 & 40 & 1000 \\ \hline
v\_1\_12\_0 & 876 & 543 & 0  & 40 & 1000 \\ \hline
v\_1\_13\_0 & 835 & 503 & 0  & 40 & 1000 \\ \hline
v\_2\_0\_0 & 738 & 441 & 36 & 40 & 1000 \\ \hline
v\_2\_2\_0 & 680 & 375 & 0  & N/A & 1000 \\ \hline
\end{tabular}}
\label{tab:pytorch_data_expanded}
\end{table}

\subsection{PyTorch Documentation Data}
\begin{table}[H]
\centering
\caption{Torch version changes over time.}
\resizebox{\columnwidth}{!}{
\begin{tabular}{|c|c|c|c|c|c|}
\hline
\textbf{Torch Version} & \textbf{Total} & \textbf{New} & \textbf{Deleted} & \textbf{Modified} & \textbf{Consistent} \\ \hline
0.4.0  & 1187 & 0   & 8   & 172 & 1179 \\ \hline
1.1.0  & 1518 & 339 & 44  & 198 & 963  \\ \hline
1.2.0  & 1548 & 74  & 10  & 43  & 1438 \\ \hline
1.4.0  & 1752 & 214 & 14  & 225 & 1507 \\ \hline
1.6.0  & 2031 & 293 & 48  & 523 & 1482 \\ \hline
1.8.0  & 2455 & 472 & 77  & 466 & 1591 \\ \hline
1.10.0 & 3324 & 946 & 596 & 237 & 1631 \\ \hline
1.12.0 & 3625 & 353 & 135 & 126 & 3051 \\ \hline
1.13.0 & 3784 & 294 & 113 & 120 & 3337 \\ \hline
2.0.0  & 4018 & 347 & 87  & 148 & 3504 \\ \hline
2.2.0  & 4187 & 256 & 0   & 68  & 3863 \\ \hline
\end{tabular}}
\label{tab:torch_version_data}
\end{table}

\subsection{Matplotlib Documentation Data}
\begin{table}[H]
\centering
\caption{API changes across versions.}
\resizebox{\columnwidth}{!}{
\begin{tabular}{|c|c|c|c|c|}
\hline
\textbf{Version} & \textbf{New} & \textbf{Deleted} & \textbf{Modified} & \textbf{Consistent} \\ \hline
3.0.3 & 0   & 180 & 135 & 4141 \\ \hline
3.2.0 & 1191 & 249 & 152 & 3875 \\ \hline
3.3.4 & 705  & 909 & 69  & 4240 \\ \hline
3.5.2 & 1505 & 523 & 84  & 4407 \\ \hline
3.6.3 & 478  & 292 & 133 & 5571 \\ \hline
3.8.3 & 631  & 0   & 59  & 6123 \\ \hline
\end{tabular}}
\label{tab:api_changes}
\end{table}

\section{Visualizing prompts given to CodeLMs in \schemelong}
\begin{figure}[H]
  \centering
  \subfloat[Prompt in a zero-shot setting]{\includegraphics[width=\linewidth]{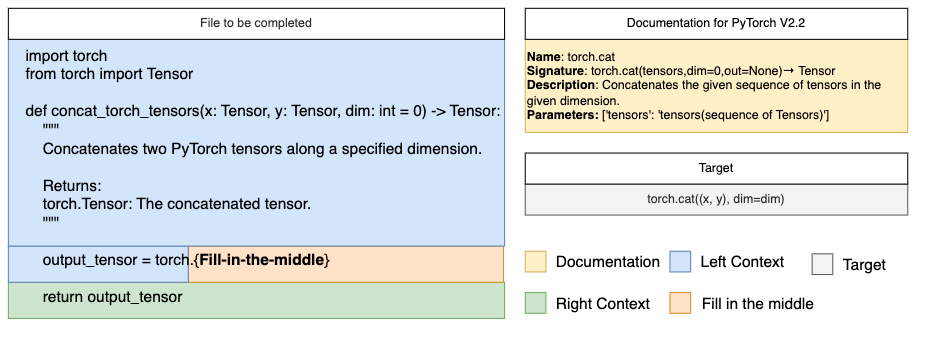}\label{fig:prompt_zero_shot}}\\
  \subfloat[Prompt when augmentation is provided]{\includegraphics[width=\linewidth]{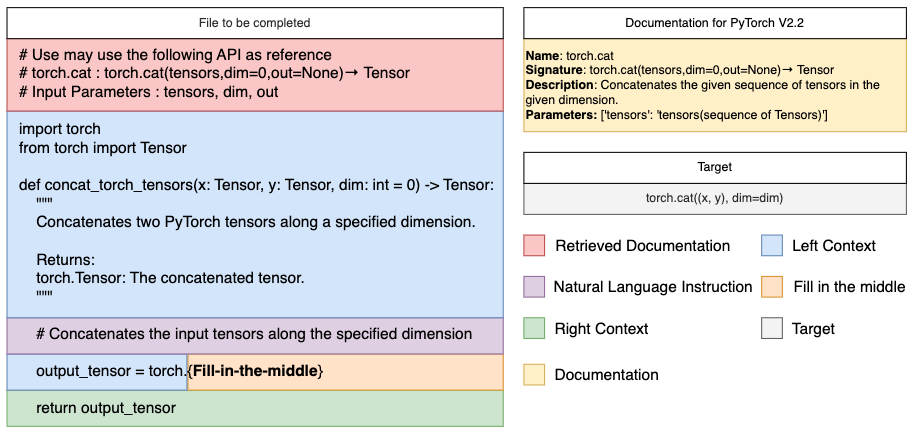}\label{fig:prompt_nl_doc}}
  \caption{Visualizing prompts given to code LLMs in \schemelong.}
  \label{fig:abc}
\end{figure}


\end{document}